\documentclass[12pt,preprint]{aastex}
\usepackage{amsmath,amssymb,bm}

\begin{document}
\label{firstpage}

\title{THE EVOLUTION OF PLANET--DISK SYSTEMS THAT ARE MILDLY INCLINED TO THE ORBIT OF A BINARY COMPANION}

\author{Stephen H. Lubow\altaffilmark{1}} \author{Rebecca
  G. Martin\altaffilmark{2}} \affil{\altaffilmark{1}Space Telescope
  Science Institute, 3700 San Martin Drive, Baltimore, MD 21218, USA }
\affil{\altaffilmark{2}Department of Physics and Astronomy, University
  of Nevada, 4505 South Maryland Parkway, Las Vegas, NV 89154, USA}

\begin{abstract}
We  determine the evolution of a
 giant planet--disk
system that orbits a member of a binary star system and is 
mildly inclined with respect to the binary orbital plane.  The planet  orbit and  disk  are initially mutually
coplanar.  
 We analyze the
evolution of the planet and the disk by analytic means and hydrodynamic
simulations.  We generally find that the planet and
the disk do not remain coplanar unless the disk mass is very large
 or the gap that separates the planet from the disk is very small.  The relative planet--disk tilt undergoes secular oscillations whose initial amplitudes are typically of order
the initial disk tilt relative to the binary  orbital plane
for disk masses $\sim$ 1\% of the binary mass or less.
The effects of a secular resonance and the disk tilt decay  
enhance the planet--disk misalignment.  
The secular resonance plays an important role for disk masses greater than the planet mass.
At later times, the accretion of disk gas by the planet causes
its orbit to evolve towards alignment, if the disk mass is sufficiently large.
The results have several implications for
the evolution of massive planets in binary systems.
\end{abstract}


\section{Introduction}
\label{intro}

Disks around young stars in binary systems may be misaligned with
respect to their binary orbital planes.  A misaligned disk in a binary
system is expected to evolve towards coplanarity due to tidal
dissipation associated with turbulent viscosity \citep{PT1995,
  Bateetal2000, LO2000, Kingetal2013}. The alignment process may occur
on relatively short timescales in binaries whose separations are small and thus
the tidal torques and associated dissipation are strong. The alignment
may also be a consequence of initial conditions in the binary formation process.
Observations show that the stellar
rotation and binary orbital axes are better aligned in closer systems
with binary separations $\la 40\,\rm  AU$ \citep{Hale1994}.  These observations provide
indirect evidence for binary--disk alignment during the star formation stage in closer binaries and misalignment
for wider binaries.

There is some direct observational evidence of disk misalignment with
respect to the binary orbital planes in wider binary systems
\citep[e.g.,][]{Jensenetal2004,Skemer2008,Roccatagliata2011}.  For the case
of the young binary system HK Tau with a projected separation of about 350
AU, circumstellar disks are observed around each component with one
disk edge--on and the other more face--on \citep{Stapelfeldt1998}.
Although the orbit of the binary is not known, at least one of the
disks must be substantially misaligned to the binary orbital plane.
Recent ALMA observations of this system by \cite{Jensen2014} suggest
that the misalignment between the two disks is $60^\circ-68^\circ$.
\cite{Williams2014} observed a wide binary (with a projected
separation of $\sim$ 440 AU) in Orion and found the misalignment
between the projected disk rotation axes to be about $72^\circ$.
These results suggest that wide binary star systems do not form
directly from a single large corotating primordial structure.
Instead, they may be subject to smaller scale effects, such as
turbulence \citep{Offner2010,Tokuda2014,Bate2012} that could result in
a lack of correlation between the rotational axes of the accreting gas
associated with the two stars.

Less direct evidence of noncoplanarity comes from the existence of
extrasolar planets whose orbits are tilted with respect to the spin
axis of the central star \citep[e.g.][]{Albrechtetal2012,
  Huber2013,Lund2014,Winn2015}.  If such planets reside in binary star
systems, these observations suggest that the planets may have formed
in disks that are misaligned with the binary orbital plane
\citep[e.g.,][]{Bateetal2010,Batygin2011, Batygin2012}.

A misaligned disk will undergo nodal precession due to the torques
caused by the companion star.  For typical protostellar disk
parameters, the disk remains nearly flat and undergoes little warping
as it precesses about the binary orbital axis \citep{Larwoodetal1996}.
A misaligned disk whose inclination angle with respect to the
binary orbital plane is between about $45^\circ$ and $135^\circ$
 can additionally undergo Kozai--Lidov oscillations 
 \citep{Martinetal2014b, Fu2015}. These oscillations cause the
disk inclination and eccentricity to vary in time. In this paper we restrict our
attention to cases where the disk inclination angle is sufficiently
small that these oscillations do not occur and the disk remains
circular.

We analyze the orbital evolution of a giant planet that interacts with
the gaseous disk in which it forms.  Such a planet will open a gap in
a disk due to tidal forces and will undergo migration as the gas
accretes towards the central star \citep{LP1986}.  The accretion is
driven by viscous forces in the disk. The planet orbit and disk are
taken to be initially coplanar and slightly misaligned with the
respect to the binary orbital plane.  In this paper, we explore the evolution of the
planet and disk.  A study along these lines was recently
carried out by \cite{XiangGruess2014} who concluded that coplanarity
between the orbital planes of the planet and the disk can be maintained.
Very recently, \cite{Picogna2015} reported results on SPH simulations with the same initial conditions that ran over longer
timescales and with higher resolution. They 
found that coplanarity is not maintained.

A giant planet--disk system that orbits a single star (not in a binary) is subject
to effects of misalignment due to mean motion resonances \citep{Borderies1984, Lubow1992}.
The misalignment can be suppressed by damping effects associated with disk viscosity \citep{Lubow2001}.
In this paper, we concentrate on the effects of the binary companion that can bring about
misalignment.

For sufficiently low mass disks, we expect that
planet--disk coplanarity cannot be maintained, since the precessional torque on the planet caused by the disk
is weak compared to the precessional torque on the disk provided by the binary companion. The disk and planet could then precess nearly independently.
On the other hand, for sufficiently high mass disks, we would expect that
coplanarity can be maintained, since the disk torque on the planet 
dominates.  We explore 
a range of disk parameters in order to understand the conditions under which
coplanarity breaks down.  We follow the tilt 
evolution in time by means of an analytic model of the secular
evolution  and by SPH simulations. 

In Section~\ref{secular} we consider analytically the secular
evolution of a planet and a rigid nonviscous outer disk that are
initially on mutually coplanar orbits, but misaligned with respect to
the orbit of a binary companion. As a test of the secular theory, in
Section~\ref{tp}, we consider a binary system with two planets
orbiting one of the stars and compare the analytic secular evolution
to that of 4--body simulations. In Section~\ref{sec:sph} we describe the results
of 3D smoothed particle hydrodynamic (SPH) simulations that model a viscous disk that
interacts with a planet in a binary system. The planet and disk are initially mutually
coplanar, but misaligned with respect to the binary orbital plane.  We
compare the results with the secular theory of nonviscous disks.
Section~\ref{disc} contains a discussion and Section~\ref{sum}
summarizes our results.

\section{Secular Evolution of Planet--Disk--Binary Systems}
\label{secular}

\begin{figure*}
\begin{center}
\includegraphics[width=8.0cm]{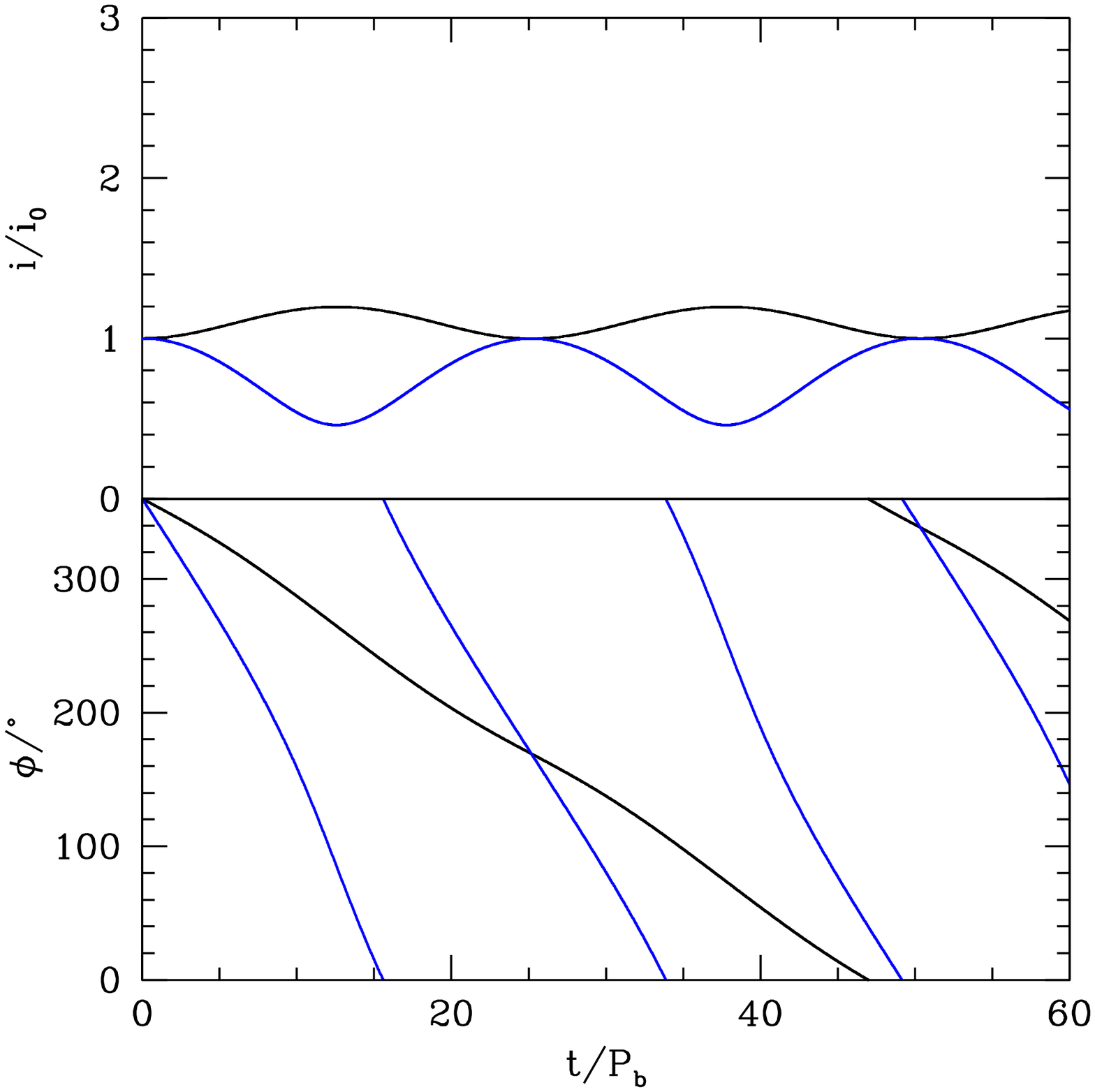}
\includegraphics[width=8.0cm]{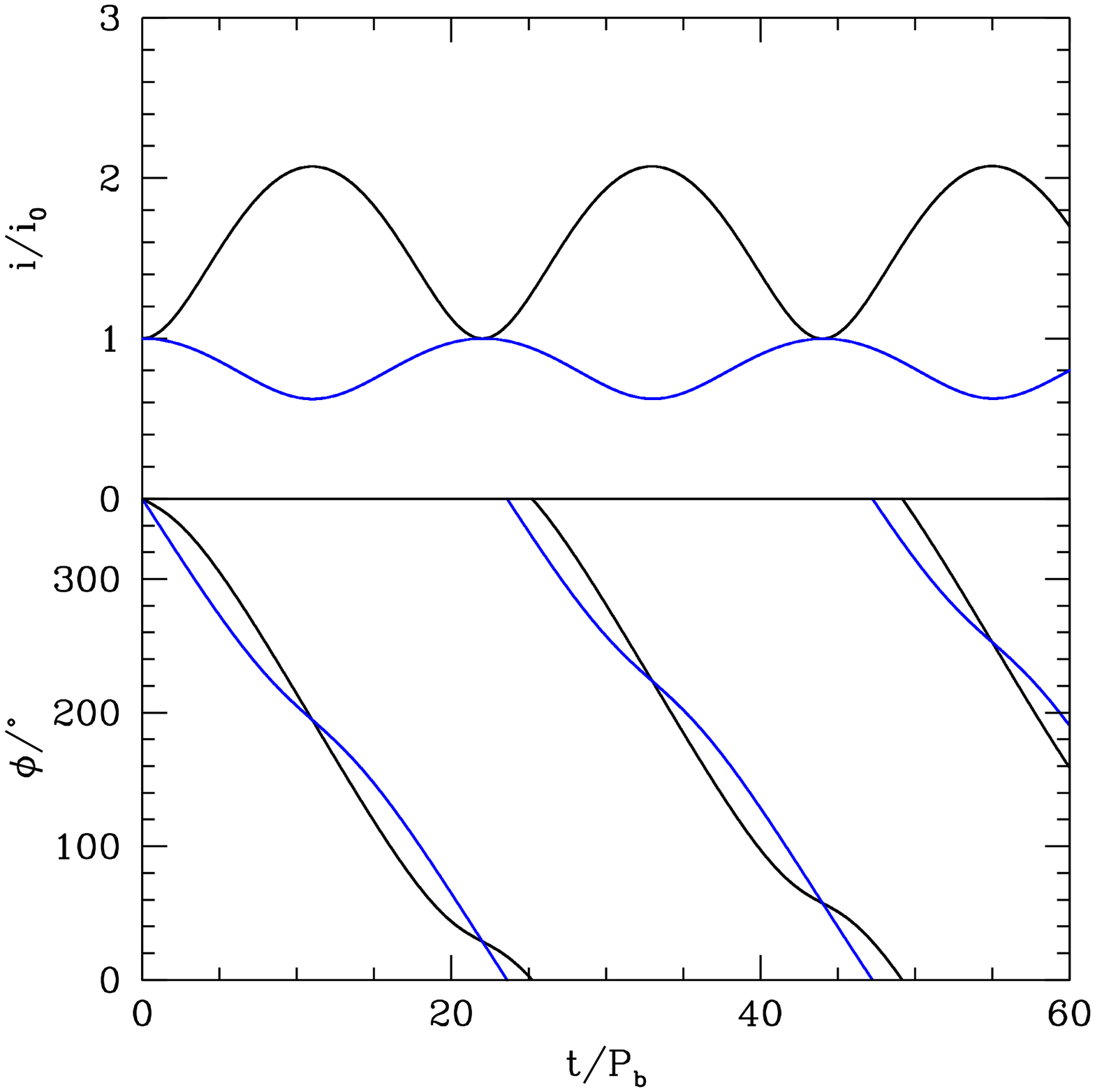}
\end{center}
\caption{Secular tilt evolution of a planet--disk system that orbits around one
  member of a binary with parameters described
  in Section \ref{params}. The time is in units of the binary orbital period $P_{\rm b}$.
   Initially, the planet orbit and disk are coplanar,
  but misaligned by angle $i_0$ from the binary orbit plane. Upper
  panels plot the tilt evolution and the lower panels plot the phase
  angle evolution. The black lines plot the planet inclination evolution, and the blue lines plot the 
  disk inclination evolution for a disk with mass $4 \times
  10^{-4}\,M$ (left) and $4 \times 10^{-3}\,M$ (right), where $M$ is the mass of the binary.}
\label{planetdisc}
\end{figure*}

\subsection{Secular Equations}
\label{sec:seceq}
In this section we consider the secular evolution of a planet and a
disk that orbit one member of a binary system.  In particular, we model a system that orbits a
central star consisting of a planet, a disk that lies exterior to the
planet, and a companion star. The orbital planes of
the planet and the disk are initially aligned, but misaligned with respect to the
binary orbital plane.   The planet, the disk, and the binary 
interact through gravitational forces.  The disk interior to
  the planet provides little torque on the other components in our SPH simulations described in Section \ref{sec:sph} and 
  we ignore its effects here. 
  The changes in the orbit of the binary due to its interactions with the planet and disk are expected to be small, since binary angular momentum is much
  larger than the angular momentum of the planet or disk.
  The binary orbit is then taken to be fixed. In addition, we assume  the binary orbit is circular and Keplerian.
  The disk is taken to be nonviscous
(nondissipative), rigid, and flat (does not warp).  

Consider a
Cartesian coordinate system $(x, y, z)$ in the inertial frame centered on  star 1 with the $z=0$ plane defined as
the binary orbital plane. The $x-$axis is parallel to the line joining the stars at the initial time. We describe the
time dependent tilt of the planet and disk relative to the  $z=0$ plane with the complex variable notation $W(t)=\ell_{\rm x}(t)+i\,
\ell_{\rm y}(t)$, where the unit angular momentum vector is denoted by
\boldmath$\ell$\unboldmath$
=(\ell_x(t),\ell_y(t),\ell_z(t))$ and $t$ is the time. 
We assume that the tilts are small, $|W| \ll 1$.

The angular frequency $\Omega$ is assumed to be Keplerian for the planet and disk,
so that $\Omega(R) =\sqrt{G M_1/R^3}$, where $M_1$ is the mass of the star 1 that lies at the disk center 
and $R$ is the distance from the center of star 1.  
 The disk extends from a radius $R_{\rm in}$ out to $R_{\rm
  out}$ and has angular momentum
\begin{equation}
J_{\rm d}=2\pi \int_{R_{\rm in}}^{R_{\rm out}}\Sigma(R) R^3 \Omega(R) \, dR,
\label{Jde}
\end{equation}
where $\Sigma(R)$ is the surface density distribution of the disk.  
The angular momentum of the planet with mass $M_{\rm p}$ and orbital
radius from the primary $a_{\rm p}$ is given by
\begin{equation}
J_{\rm p}=M_{\rm p}a_{\rm p}^2 \Omega(a_{\rm p}).
\label{Jpe}
\end{equation}
The binary
companion star 2 has mass $M_2$ and orbital radius $a$.

We apply the secular evolution equations for the gravitational 
interactions between slightly misaligned components in 
\cite{Lubow2001}. 
 In this model, the torques between the components of the system (planet, disk, and binary) are evaluated
in the small angle approximation for their relative tilts. 
In Section~\ref{tp}, we consider a binary system with two planets
orbiting one of the stars and compare the analytic secular evolution
to that of 4--body simulations. We find that the analytic results are accurate
for initial tilt angles up to about $20^\circ$,  due in part to the small angle approximation that
limits the accuracy at larger initial tilt angles.
In addition, the components
are assumed to remain on circular orbits. Consequently, this formalism
cannot be used to analyze planet-disk systems undergoing Kozai-Lidov oscillations,
where the relative tilts are large and the planet orbit and disk may acquire
substantial eccentricity.

The interaction between the components $j$ and $k$ 
is described  by a linear model through coupling coefficients denoted by $C_{jk}$.
The evolution equations for the planet tilt $W_{\rm p}(t)$ and the disk tilt $W_{\rm d}(t)$ are given by 
\begin{equation}
J_{\rm p} \frac{d W_{\rm p}}{dt}=  i C_{\rm pd}(W_{\rm d} -W_{\rm p})  - i C_{\rm ps} W_{\rm p}
\label{Wp}
\end{equation}
and
\begin{equation}
J_{\rm d} \frac{d W_{\rm d}}{dt}=  i C_{\rm pd}(W_{\rm p} -W_{\rm d})  - i C_{\rm ds} W_{\rm d},
\label{Wd}
\end{equation}
where subscripts ${\rm p},$ ${\rm d}$, and ${\rm s}$  refer to the planet, disk and companion star 2, respectively. 
The first term on the right-hand side of Equation (\ref{Wp}) is the horizontal torque (along the plane of the binary)
on the planet due to the disk and the second term is the  horizontal torque on the planet due to the binary companion  star.
The first term on the right-hand side of Equation (\ref{Wd}) is the horizontal torque 
on the disk due to the planet and the second term is the  horizontal torque on the disk due to the binary companion.

The coupling coefficients are given by
\begin{equation}
C_{\rm pd}=2\pi \int_{R_{\rm in}}^{R_{\rm out}} G M_{\rm p} R \, \Sigma(R) \, K(R,a_{\rm p})\,dR,
\label{Cpd}
\end{equation}
\begin{equation}
C_{\rm d s}=2\pi \int_{R_{\rm in}}^{R_{\rm out}} G M_2 R \, \Sigma(R) \, K(R,a)\,dR,
\end{equation}
and
\begin{equation}
C_{\rm p s}=GM_{\rm p}M_2K(a_{\rm p},a),
\end{equation}
where the symmetric kernel, with units of inverse length, is given by
\begin{equation}
K(R_j,R_k)=\frac{R_j R_k}{4\pi} \int_{0}^{2\pi} \frac{ \cos \phi \,d\phi}{(R_j^2+R_k^2-2R_j R_k\cos \phi)^{3/2}}.
\label{K}
\end{equation}
The kernel contains a singularity as $|R_j - R_k| \rightarrow 0$. This singularity is resolved by
the finite thickness of the disk $H$. We assume in this paper that the separation
between the components is greater than $H$ and consequently we do not smooth the kernel.

Consider a time-periodic vector  of tilts with frequency $\omega$
\begin{equation}
{\bf W}(t)=  {\bf \tilde{W}} \exp{(i \omega t)}
\end{equation}
where
\begin{equation}
{\bf \tilde{W}} =
\begin{pmatrix} 
\tilde{W}_{\rm p} \\ \tilde{W}_{\rm d}\\
\end{pmatrix}.
\end{equation}
The tilts satisfy the matrix equation
\begin{equation}
M {\bf \tilde{W}}={\bf 0},
\label{matrix}
\end{equation}
where 
\begin{equation}
M=
\begin{pmatrix}
\omega J_{\rm p} +C_{\rm pd}+C_{\rm ps}  & -C_{\rm pd}   \\
-C_{\rm pd}  & \omega J_{\rm d}+C_{\rm pd}+C_{\rm d s}    \\
\end{pmatrix}.
\end{equation}
The frequency $\omega$ can be shown to be real.

Although we assume that the actual tilts are small, for simplicity in this linear problem, we set $\tilde{W}_{\rm p}=1$ to solve for the eigenvectors and the eigenfrequencies. The contributions of these eigenvectors to the tilt evolution are determined by initial
conditions. 
We denote the two eigenvectors by  ${\bf \tilde{W}}_j$ for $j=1$ and 2.
We have determined these eigensolutions analytically.
We apply these analytic solutions to determine the numerical results
that we describe below.

With these two eigensolutions for the matrix equation, we can solve any initial value problem.  We determine the contributions of the eigenvectors  ${\bf \tilde{W}}_i$ through
equation
\begin{equation}
c_1 {\bf \tilde{W}}_1 +c_2 {\bf \tilde{W}}_2 =   {\bf \tilde{W}}_0
\end{equation}
by solving for constants $c_1$ and $c_2$, where $ {\bf \tilde{W}}_0= (W_{\rm p}(0), W_{\rm d}(0))^T$ and
$W_{\rm p}(0)$ and $W_{\rm d}(0)$ are the initial tilts for the orbits of the planet and disk, respectively,
and where $T$ denotes the transpose of the vector from row to column  form. 
 The evolution of the tilts of the planet-disk components is then given by
\begin{equation}
{\bf{W}}(t)=c_1 {\bf \tilde{W}}_1 e^{i \omega_1 t}+c_2 {\bf \tilde{W}}_2 e^{i \omega_2 t},
\label{Wt}
\end{equation}
where ${\bf{W}}(t) = (W_{\rm p}(t), W_{\rm d}(t))^T$ describes the tilt evolution
of the planet and disk, respectively. 

We consider systems with the initial tilts given by ${\bf \tilde{W}}_0=(
i_0,  i_0)^T.$
This vector represents a planet and a disk
whose orbital planes are initially aligned with each other, but
are slightly misaligned with respect to the orbital plane of the binary with small $i_0$.
The inclination of each component $i_j(t)$ relative to the binary orbital plane is given by 
\begin{equation}
i_j(t)=  |W_j(t)|,
\label{iW}
\end{equation}
for $j={\rm p}, {\rm d}$.  The linear model assumes that tilts $i_{\rm p}(t)$ and $ i_{\rm d}(t)$ are small.
The phase angle is given by 
\begin{equation}
\phi_j(t) =
\tan^{-1}\left(\frac{Im(W_j(t))}{Re(W_j(t))}\right).
\end{equation}
We apply these equations in the following subsections to determine the evolution
of some planet-disk-binary systems.

\subsection{System Parameters}
\label{params}

\begin{figure*}
\begin{center}
\includegraphics[width=8.0cm]{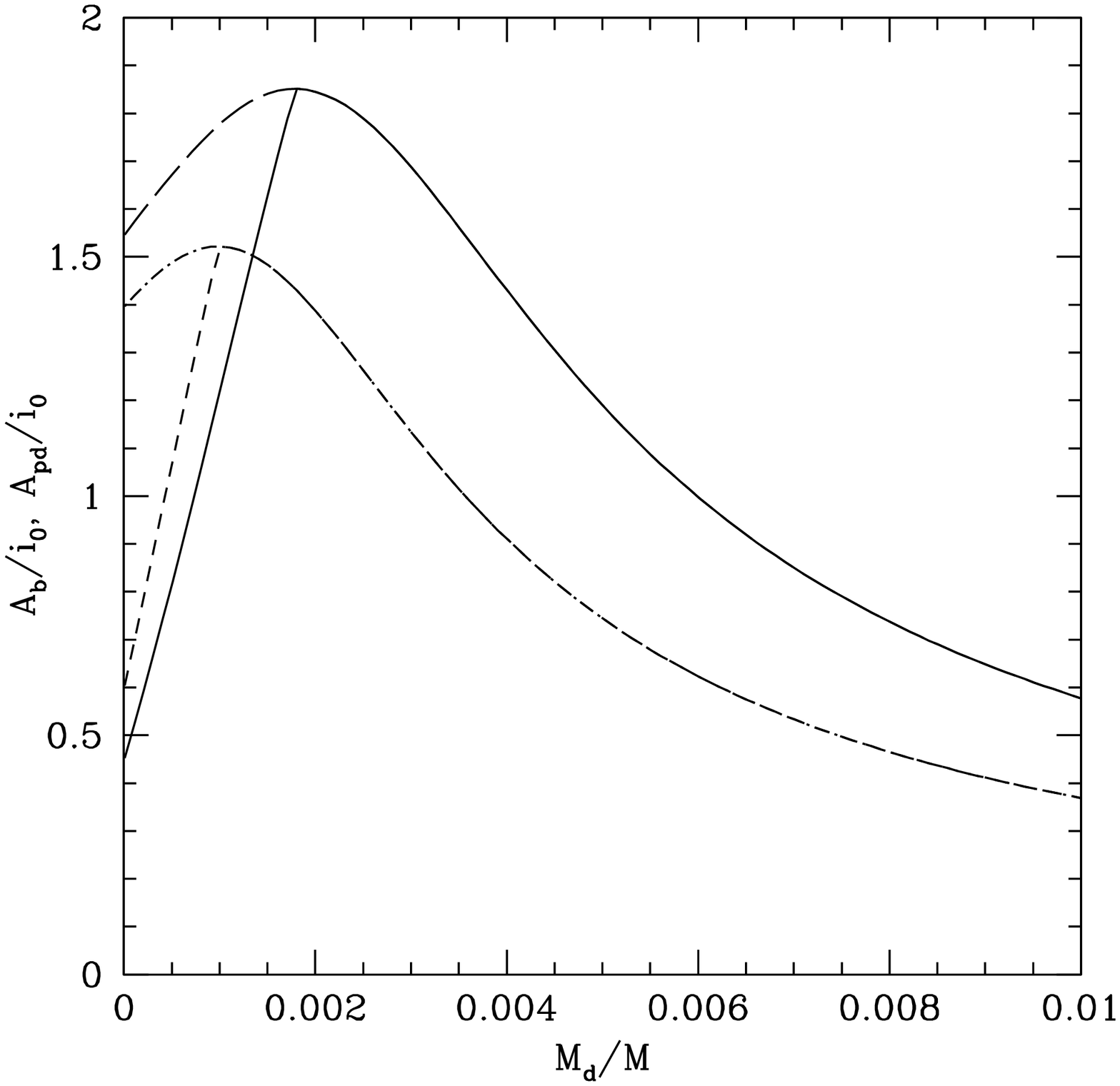}
\includegraphics[width=8.0cm]{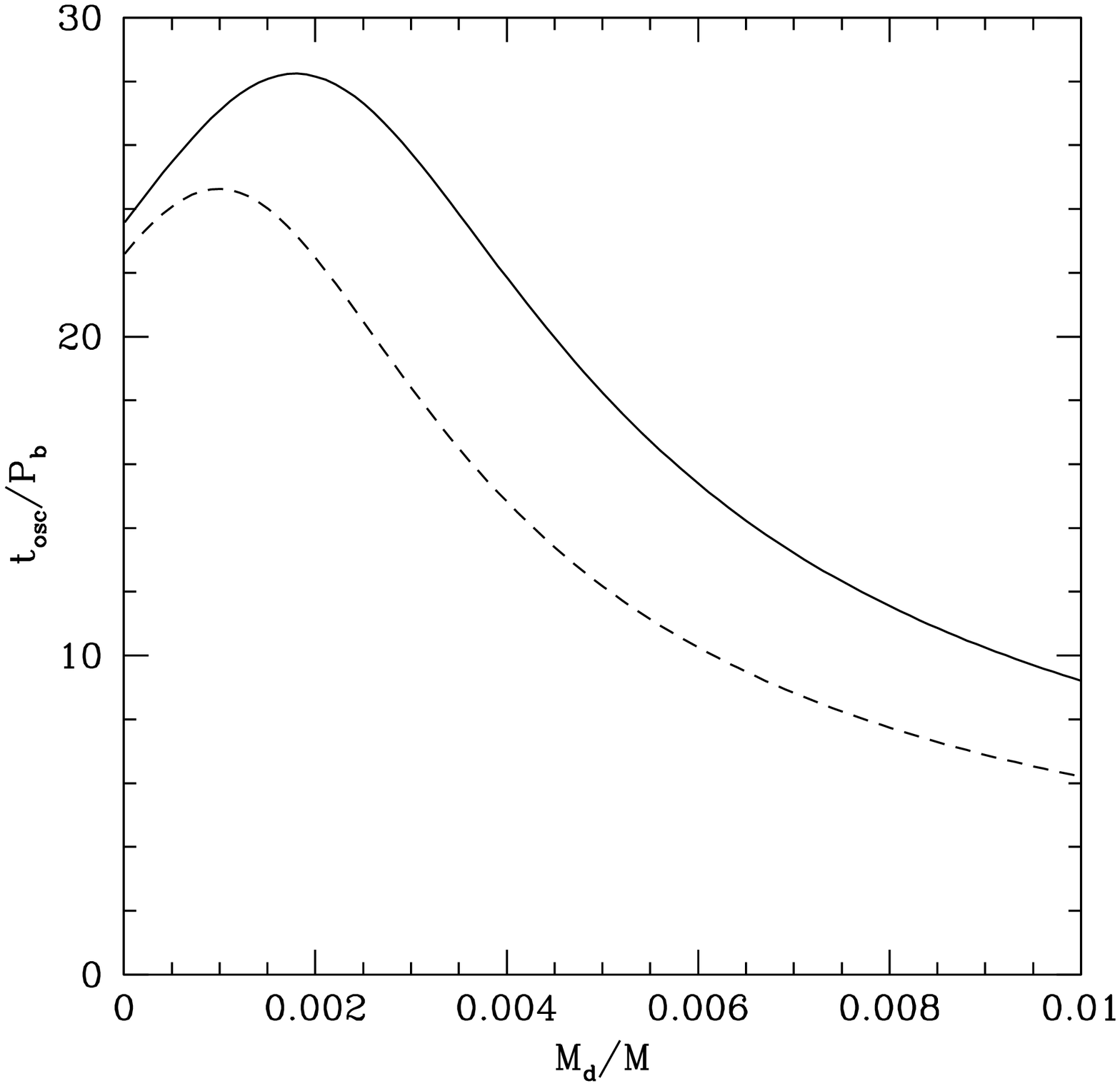}
\end{center}
\caption{Left: Short--dashed (solid) line plots 
 the tilt oscillation amplitude $A_{\rm b}$ defined by Equation (\ref{Ab}) as a function of disk mass, normalized 
 by binary mass $M$, with disk
  inner radius $0.13\,a$ ($0.14 \, a$). The planet has an orbital
  radius of $0.1\,a$ and a mass of $1 \times 10^{-3} M$.  The dashed--dotted (long--dashed) line plots the
  the tilt oscillation amplitude $A_{\rm pd}$ defined by Equation (\ref{Apd}) as a function of the
  mass of a disk whose inner radius is $0.13\,a$ ($0.14 \, a$).  The planet and the disk
  orbital planes are initially coplanar, but tilted by angle $i_0$ with
  respect to the binary.  Note
  that the solid and long--dashed, as well as the short--dashed and
  dashed--dotted, overlap at larger disk masses (beyond the plot peak
  values) because the mean precession rates of the planet and disk are locked,
  although significant relative tilts are present.  Right: The
  inclination oscillation periods for the solid and short--dashed line
  cases plotted in the left panel. }
 \label{deltaimass}
\end{figure*}

We choose a set of standard system parameters and consider 
the effects of variations from these values.
The parameter values for the disk are selected as plausible 
and are not tuned to quantitatively match the results of the SPH simulations
that are described in Section \ref{sec:sph}. The results of the secular models
demonstrate trends that explain properties of
the simulations that are due to gravitational torques.

The surface density of the disk is assumed to follow
$\Sigma(R)=\Sigma_0 (a/R)^{3/2}$, where $\Sigma_0$ is a constant
defined through the constraint that the mass of the disk, $M_{\rm d}$, is
given.  The disk is assumed to extend only exterior to the orbit of
the planet. The standard disk mass is taken to be $4 \times 10^{-3} M$, 
where $M$ is the binary mass.

  As discussed in Section \ref{sec:seceq},
  the binary orbit is taken to be circular.
 The value of the disk outer radius $R_{\rm out}$ is taken to be $0.25 a$
that is about equal to the tidal truncation
radius of a disk in a coplanar equal mass binary system 
\citep[see][]{Paczynski1977}.  The tidal truncation radius for a noncoplanar configuration
may be somewhat larger due to the weakening of the tidal torques \citep{Lubowetal2015}.
The planet is taken to have mass $M_{\rm p}=1 \times 10^{-3}\,M$
and orbital radius $a_{\rm p}=0.1\,a$.
The planet and disk orbit star 1 in an equal mass binary with $M_1=M_2=0.5\,M$.

The clearance between the orbit of planet and the
inner edge of the disk is determined by the size of the gap. The gap
size depends on the level of disk viscosity, planet mass, etc. Also,
the "gap" region is not completely clear of gas. 
Using the  gap density profile $\Sigma(r)$ corresponding to \cite{Bateetal2003}
 for a planet--to--star mass ratio of $1 \times 10^{-3}$ (a $1 M_{\rm J}$ planet orbiting a $1 M_{\odot}$ star),
 we compute a value of $C_{\rm pd}$ in Equation (\ref{Cpd}) with $R_{\rm in}= a_{\rm p}$. 
 We then determine the effective disk inner radius $R_{\rm in}$ that gives the same value of $C_{pd}$, but 
for a disk with an empty gap region and a sharp disk inner edge. This procedure yields in an effective disk inner radius
  $R_{\rm in}= 1.3 a_{\rm p}$
 For a planet--to--star ratio of $2 \times 10^{-3}$, as we consider here ($M_{\rm p}= 1 \times 10^{-3} M$), the gap size is expected to be 
 somewhat larger, perhaps by a factor of $2^{1/3}$, based on scaling by the Hill radius, 
 and we adopt a fiducial value of $R_{\rm in}= 1.4 a_{\rm p}$.

\subsection{Effect of Varying the Disk Mass}

Fig.~\ref{planetdisc} shows the evolution of the inclinations and phase angles
in a system with the standard parameters described in Section \ref{params},
except that we consider variations to the disk mass $M_{\rm d}$. Since the secular
evolution model is linear, the tilts scale with $i_0$, and so we plot $i(t)/i_0$.
 
As seen in Fig.~\ref{planetdisc}, the inclinations of the planet and the disk for
the smaller disk mass of $4\times 10^{-4} M$ oscillate away from each
other, while each precesses on different timescales. The planet
precesses more slowly than the disk because it is farther away from
the companion star.  For the larger disk mass of $4\times 10^{-3} M$,
the precession of the planet is clearly affected by its interaction
with the disk that causes its precession rate to be close to that of
the disk that is somewhat longer than in the left panel.  For both disk masses, the inclination of the planet is generally
larger than its initial value, while the inclination of the
disk is generally smaller than its initial value. The planet and disk spend very
little time in a coplanar configuration. Instead, we find generally that $i_{\rm p}(t)-i_{\rm d}(t)~ \sim i_0$.

The level of planet--disk misalignment undergoes periodic oscillations.
We measure this misalignment by the amplitude of these oscillations.
The left hand panel of Fig.~\ref{deltaimass} plots the normalized oscillation 
amplitudes of planet--disk  tilt differences, as a function of disk
mass. The right hand panel plots the oscillation period as a function of disk mass. The
amplitude is determined in two different ways. One set of
lines (solid and short--dashed) plots the oscillation
 amplitude (the maximum value over time) of the difference in the planet  and disk tilts relative to the binary orbital plane $A_{\rm b}$, normalized by the initial tilt $i_0$. 
The amplitude in this case is defined by
\begin{equation}
A_{\rm b}= \max_{t} \left(|W_{\rm p}(t)| - |W_{\rm d}(t)|\right).
\label{Ab}
\end{equation}
The other set of lines (long-dashed and dashed--dotted) plots the oscillation
amplitude of the planet tilt relative to the disk $A_{\rm pd}$, normalized by the initial tilt $i_0$. 
The amplitude in this case is defined by
\begin{equation}
A_{\rm pd}= \max_{t}\left|W_{\rm p}(t) - W_{\rm d}(t)\right|.
\label{Apd}
\end{equation}
This quantity depends on the
relative phasing of the planet and disk tilts.  For disk masses beyond
the peaks in the curves, both ways of measuring the tilt differences
coincide because the planet and disk mean precession rates are locked.

\begin{figure}
\begin{center}
\includegraphics[width=8.0cm]{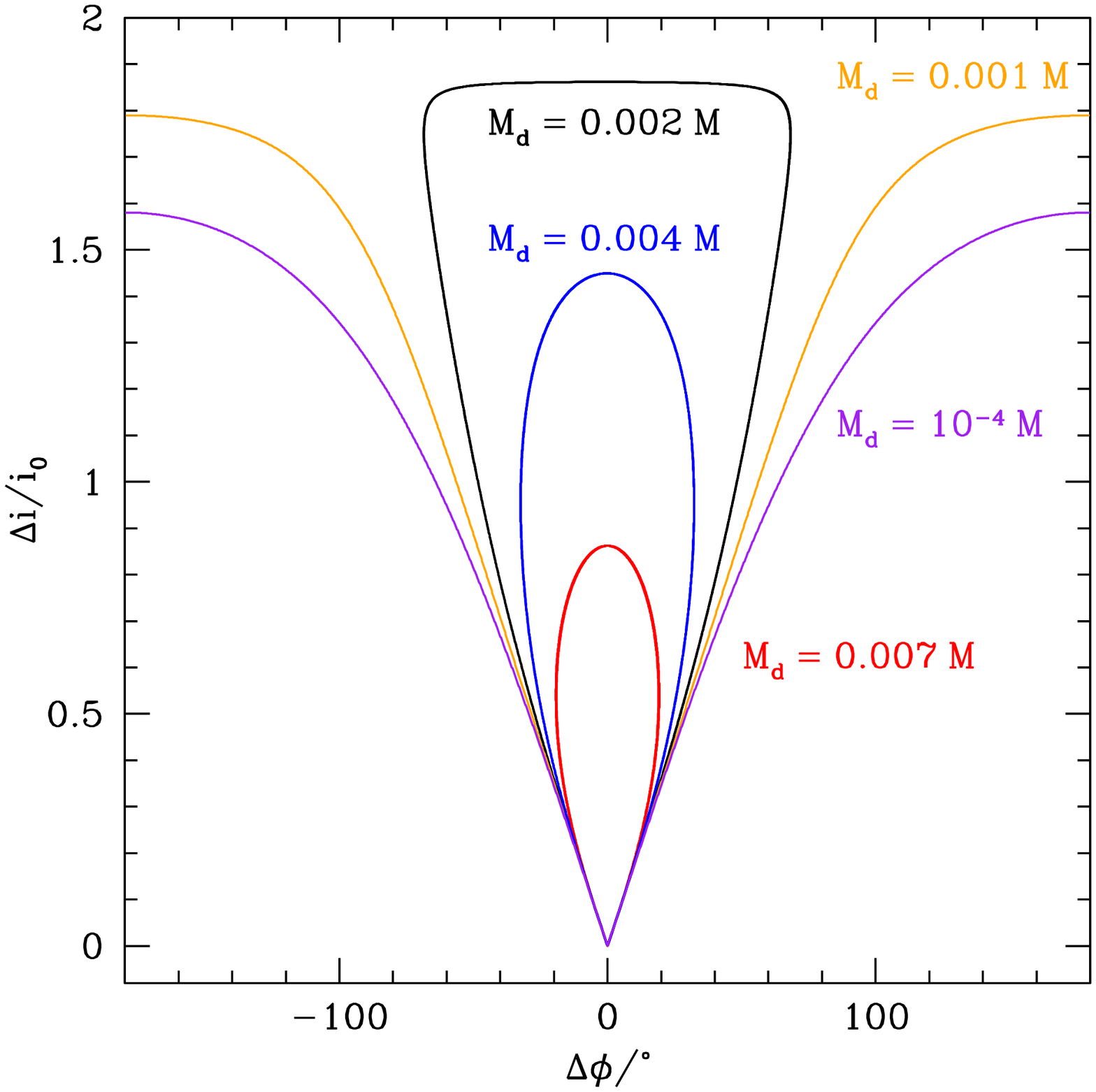}
\end{center}
\caption{Plot (phase portrait) of the relative planet--disk tilt $\Delta i/i_0= |W_{\rm p}-W_{\rm d}|$ versus nodal phase difference $\Delta \phi=\phi_{\rm p} - 
\phi_{\rm d}$ for the model plotted in Fig.~\ref{deltaimass} with disk inner radius $0.14  a$ and various disk masses external
to the orbit of the planet.}
\label{phasepor}
\end{figure}

In the left panel
of Fig.~\ref{planetdisc}, the  maximum inclination difference is $|W_{\rm p}(t)| - |W_{\rm d}(t)|=0.68 i_0$ and this occurs at time of about $t=12 \,P_{\rm b}$, corresponding to an
oscillation period of about $t_{\rm osc}=25\,P_{\rm b}$. These values correspond to the oscillation amplitude $A_{\rm b}/i_0$ and oscillation
period $t_{\rm osc}$ that lie on the solid lines in the
left and right panels of Fig.~\ref{deltaimass}, respectively, for a disk mass of $4\times 10^{-4} M$.

The combined planet-disk angular momentum is generally not conserved. 
In the left panel of Fig.~\ref{planetdisc}, the planet and disk
precess independently due to the torque caused by the binary.
While precessing at different rates, their individual angular
momentum vectors vary differently in time.  In the right panel, they precess
together while undergoing angular momentum changes due to the binary
torque.

We might expect that with increasing disk mass the planet becomes more aligned with the disk.
 However, there   are peaks of
$A_{\rm pd}/i_0$ in the left panel of Fig.~\ref{deltaimass}. 
It is somewhat surprising that the level planet-disk misalignment can increase with increasing disk mass.
In Section \ref{sec:secres} we show that this peak is associated with a secular resonance.
For a disk inner radius of $0.14 \,a$ in Fig.~\ref{planetdisc},
the peak occurs near $M_{\rm d}=  2 \times 10^{-3}\,M$.  For disk masses smaller than this value,
the planet and disk precess independently. 
The tilt oscillation frequency is equal to the difference in the nodal precession frequencies between the
  disk and planet. However, for increasing disk
masses greater than this value, the precessions of the planet and disk become locked and the
inclination differences and the oscillation periods decrease. For a smaller disk inner radius (smaller gap),
the planet and disk interact more strongly and reach peak values in Fig.~\ref{deltaimass}
at a lower disk mass. This effect is described further in Section \ref{sec:gap}. Note however that except for small values of the disk mass,
the inclination difference is generally comparable to or greater than the initial inclination,  
$\Delta i = i_{\rm p} - i_{\rm d} \ga i_0$, for disk masses up to $0.01 M$. This result suggests that departures from coplanarity
are significant  in many cases.

 Fig.~\ref{phasepor} plots the relative disk--planet tilt $\Delta i/i_0= |W_{\rm p}-W_{\rm d}|$ as a function of nodal phase difference $\Delta \phi=\phi_{\rm p} - 
\phi_{\rm d}$ for the model plotted in Fig.~\ref{deltaimass} with disk inner radius $0.14  a$ for different disk masses.
For a disk mass of $1\times10^{-4} M$, the planet is only slightly affected by the presence of the disk and they precess independently at different
rates, resulting in a  fully circulating phase difference.   
 $\Delta i$ varies in this case because the disk undergoes small
amplitude tilt oscillations due to its interaction with the planet.  When the planet and disk phase angles are
  aligned, the relative tilt is zero. When the difference in the phase
  angles is $\pm 180^\circ$, the relative tilt is maximum. 

For a larger disk mass of $1 \times 10^{-3} M$, Fig.~\ref{phasepor} shows that the interactions lead the system to undergo stronger 
tilt oscillations with larger planet--disk relative tilts, while still fully circulating in phase. When the disk mass becomes large enough for them begin to precess
together ($2 \times 10^{-3} M$), the phase difference is limited and the system is librating.  The mean  precession
rates for the planet and for the disk over a libration cycle are then equal, unlike the fully circulating case at lower disk masses.
The planet and disk mean precession rates are then locked.
At this disk mass, the amplitude of relative tilt  oscillations is maximum, as discussed above. 
At the times of
  both maximum and minimum relative tilt $\Delta i/i_0$, the disk and planet have the same phase, $\Delta \phi=0$. For increasingly larger disk
masses, there is a decrease in the amplitude of the $\Delta i$ oscillations and the range of  phase $\Delta \phi$.  However, the relative tilts can be significant compared to the initial tilt $i_0$ with respect to the binary for disk masses up to one or more  percent of the binary mass.

\subsection{Effect of Varying the Disk Inner Radius \label{sec:gap}}

For the higher disk mass especially, the evolution of the disk--planet
system is somewhat sensitive to the radius of the inner edge of the
disk for a fixed planet orbital radius. The reason is that the kernel $K$ defined Equation (\ref{K})
varies as $(R_i - R_j)^{-2}$ for $R_i \simeq R_j$, resulting in a
variation of $\sim (R_{\rm in}-a_{\rm p} )^{-1}$ in the disk--planet
coupling coefficient $C_{\rm pd}$.

\begin{figure*}
\begin{center}
\includegraphics[width=8.0cm]{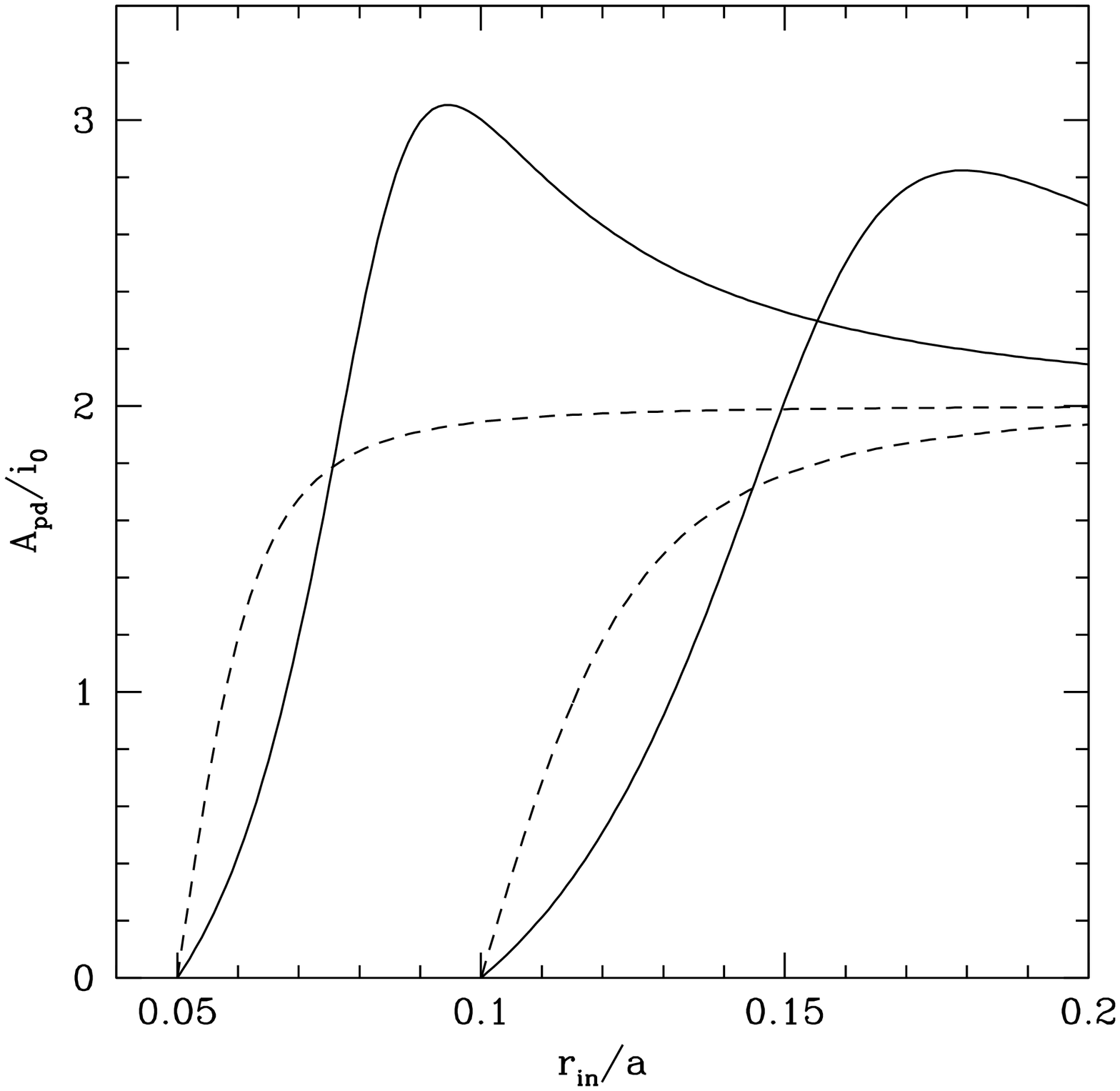}
\includegraphics[width=8.0cm]{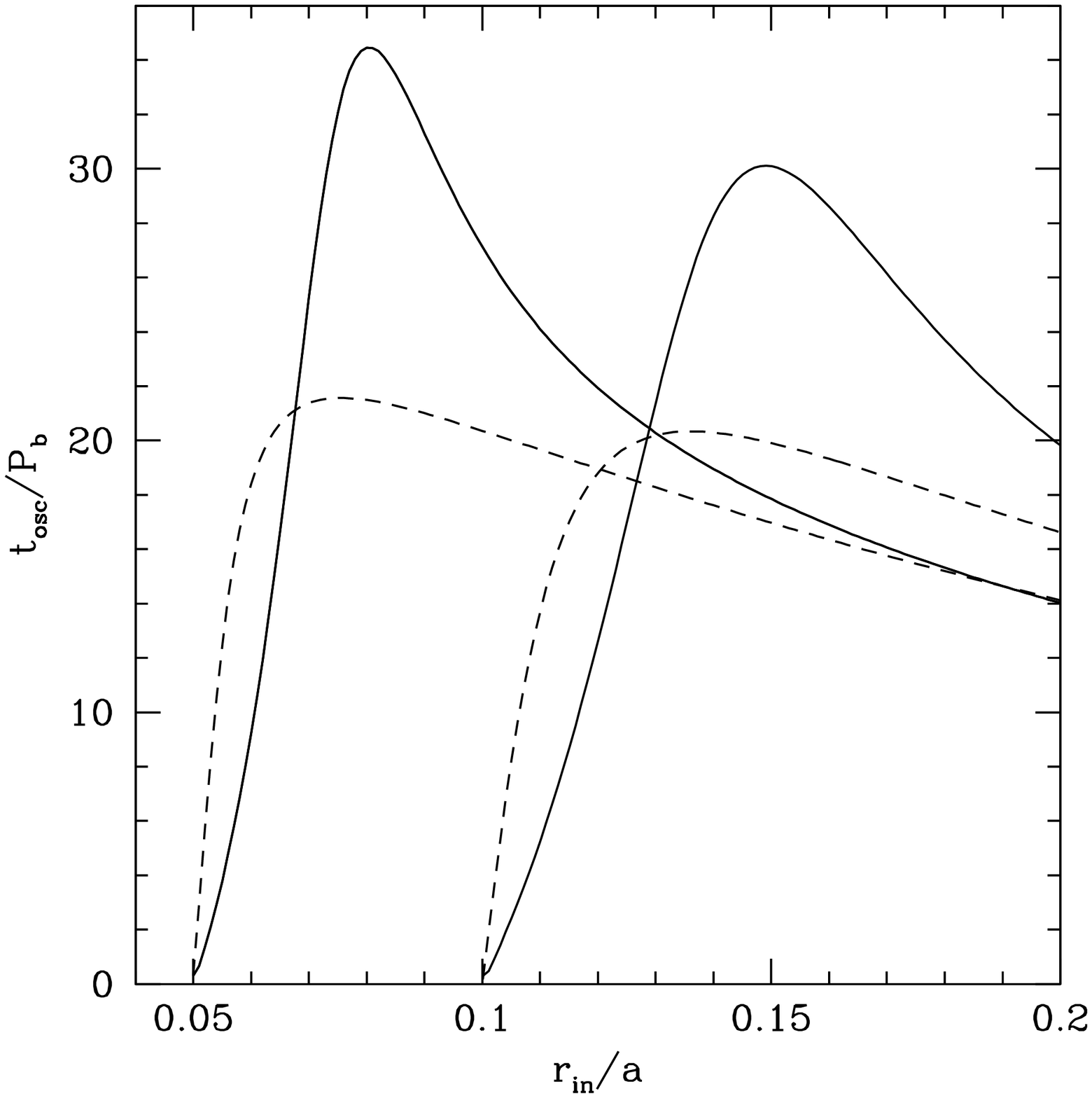}
\end{center}
\caption{The amplitude of the tilt oscillations of the planet 
  relative to the disk, $A_{\rm pd}$,  (left) and the oscillation period (right) 
  as a function of disk inner radius $r_{\rm in}/a$. The tilt is normalized by the initial planet--disk
  misalignment to the binary of $i_0$. The solid lines plot the
  higher mass disk case with $M_{\rm d}=4 \times 10^{-3}\,M$ and the
  dashed lines plot the case of a lower mass disk with $M_{\rm d}=4
  \times 10^{-4}\,M$, where $M$ is the mass of the binary.  In each panel, the left solid (dashed) line is
  for the case of a planet with orbital radius $a_{\rm p}=0.05\,a$
  interacting with the higher (lower) mass disk.  The right solid
  (dashed) line is for a planet
  with orbital radius $a_{\rm p}=0.1\,a$ interacting with the higher
  (lower) mass disk.  }
\label{deltai}
\end{figure*}

For a smaller disk inner radius (smaller gap),
the planet and disk interact more strongly and reach peak relative tilt values in Fig.~\ref{deltaimass}
at a lower disk mass. 
The left panel of Fig.~\ref{deltai} shows the amplitude of the planet--disk relative inclination oscillations,
$A_{\rm pd}/i_0$, as a function of disk inner radius with disk mass $M_{\rm
  d}=4 \times 10^{-3}\,M$. The right panel shows the oscillation
period. If the inner edge of the disk is very close to the planet,
then the oscillation amplitude may be very small. There is a peak in
the oscillation amplitude. If the disk inner radius is smaller than
the  inner radius at the peak, then the mean precession rates of the planet and the disk are locked. But for
a larger inner disk radius, they are more disconnected and precess
independently.  When the planet and disk do not interact with each other,
their maximum relative tilt is twice the initial tilt of $i_0$ that  occurs when they are separated in
phase by $180^{\circ}$. (As noted below Equation (\ref{K}), we assume that the gap size is larger than the disk thickness in evaluating
$K$. This plot is then valid for cases
where $R_{\rm in} \ga H + a_{\rm p}$.)
  But unless the gap is small, less
than about $0.2 a_{\rm p}$, the planet and a larger mass disk can be
substantially misaligned with misalignment angle greater than about half of their
initial tilts relative to the binary orbital plane.

\subsection{Effect of the Varying the Orbital Radius of the Planet}

In Fig.~\ref{deltai} we include plots of the relative disk--planet
inclination for a planet at a smaller orbital radius of $0.05\,a$ for different disk inner radii.
The planet--disk relative inclinations at the two planet orbital radii are very
similar.
This plot then shows that the orbital radius of the planet does not strongly
affect the amplitude of the oscillations.  The more important factor is
the disk gap size.

\subsection{Effect of Varying the Disk Outer Radius}

\begin{figure}
\begin{center}
\includegraphics[width=8.0cm]{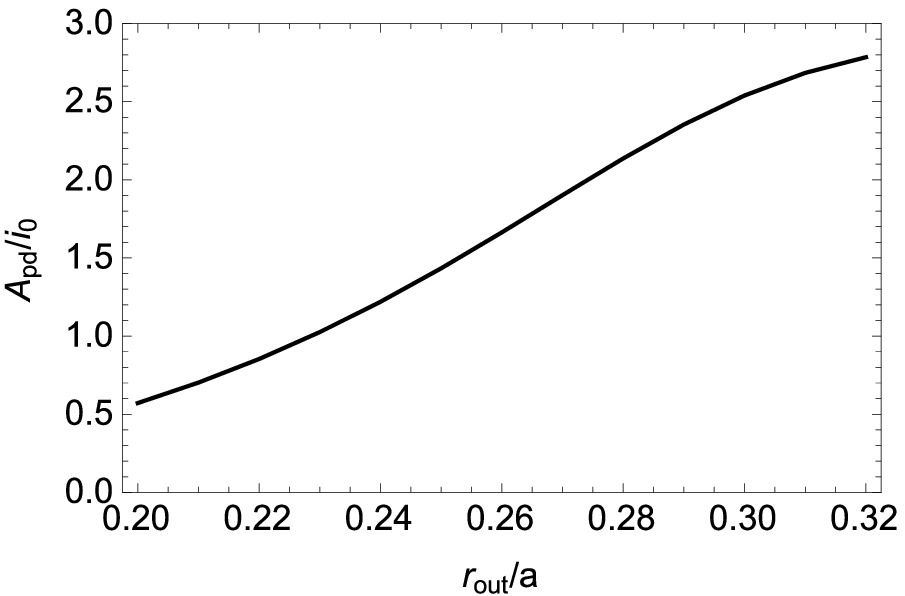}
\end{center}
\caption{The amplitude of the tilt oscillations of the planet orbit
  relative to the disk  
  as a function of disk outer radius $r_{\rm out}/a$ for the standard model.  }
\label{fig:rout}
\end{figure}

The disk outer radius is controlled by a competition between the disk viscous torques that act
to expand the outer parts of the disk with tidal torques that act to truncate the disk.
As discussed in Section \ref{params}, for the standard model we choose the outer disk radius $R_{\rm out}$ to be near the tidal truncation
radius of a disk in a coplanar binary system
\citep{Paczynski1977}.   
However, in a misaligned binary, the
disk can be somewhat larger because the tidal torque that truncates the disk
is reduced \citep{Lubowetal2015}.
In Fig.~\ref{fig:rout} we plot the amplitude of the oscillations of the planet orbital tilt
 relative to the disk
  as a function of disk outer radius for the standard model described in Section \ref{params}. 
  A larger disk  experiences a stronger binary torque that leads to greater planet--disk misalignment.
  The results show that the amplitude of the relative tilt  oscillations $A_{\rm pd}/i_0$ is a sensitive function of the disk outer radius.
  However, the relative tilts are substantial over the plotted range of disk outer radii.

\subsection{Role of Secular Resonances}
\label{sec:secres}

Secular resonances are known to play an important role in the dynamics of small mass objects, such as asteroids,
in multi-planet systems \citep{Murray1999}.
Secular resonances occur when there is a matching of the precession frequency of a test particle (e.g., an asteroid) 
with the frequency of one of the precessional modes of the planetary system. 
The frequencies involved are much lower than in the mean-motion resonance case, i.e., Lindblad resonances. 
At such a resonance, the motion of a test particle  can be strongly driven 
by the planets, resulting in a high orbital inclination (for a nodal resonance) or eccentricity 
(for an apsidal resonance). 
In the context of this paper, we are concerned with nodal secular resonances.
An example of such resonance is the so-called $\nu_{16}$ resonance  caused by
Saturn and Jupiter driving
strong vertical motions in the asteroid belt, resulting in a narrow gap at the radius where a
secular resonance condition is satisfied among the nodal precession frequencies. 

We analytically determined the properties of the peak value of $A_{\rm pd}/i_0$ as a function
of disk mass, such as is seen in the left panel of Fig.~\ref{deltaimass}. 
 Using analytic expressions for $W_{\rm p}$ and $W_{\rm d}$, we find that this
 the maximum 
  occurs when
 \begin{equation}
 \omega_{\rm d s}=  \omega_{\rm pd} \left(1+ \frac{J_{\rm p}}{J_{\rm d} } \right)+ \omega_{\rm p s},
 \label{resc}
 \end{equation}
 where $\omega_{\rm d s} = - C_{\rm d s}/J_{\rm d}$ is the precession rate of the disk
 caused by the binary companion star, $\omega_{\rm p d} = - C_{\rm p d}/J_{\rm p}$ is the precession rate of the planet
 caused by the disk, and $\omega_{\rm p s} = - C_{\rm p s}/J_{\rm d}$ is the precession rate of the planet
 caused by the binary companion star. In this equation, $\omega_{\rm pd}$ and $J_{\rm d}$
 are linear functions of disk mass $M_{\rm d}$, while the other parameters are independent
 of disk mass.
 The maximum value of the planet--disk tilt oscillation amplitude for varying  disk mass is given by
 \begin{equation}
\frac{A_{\rm max}}{i_0} = \sqrt{1+ \frac{J_{\rm d}}{J_{\rm p} } },
\label{imax}
\end{equation}
where $J_{\rm d}$ is the value of the disk angular momentum that occurs when Equation (\ref{resc})
is satisfied.

For $J_{\rm p} \ll J_{\rm d}$, the resonance full-width at half-maximum in terms of disk mass $\delta M_{\rm d}$ 
is given by
 \begin{equation}
\frac{\delta M_{\rm d}}{M_{\rm d}} = 4 \sqrt{3} \sqrt{\frac{J_{\rm p}}{J_{\rm d }}},
\label{dm}
\end{equation}
where $J_{\rm d}$ and $M_{\rm d}$ are the values of the disk angular momentum and disk mass that occurs when Equation (\ref{resc})
is satisfied. This resonance width estimate assumes that the gap size is maintained in taking this low $J_{\rm p}$ limit. While this assumption will not likely hold,
this estimate is made for comparison with the $v_{16}$ resonance, as described below.

Equation (\ref{resc}) is essentially a secular resonance condition. 
If we consider the planet to be a very small mass
object so that $J_{\rm p} \ll J_{\rm d}$, then Equation (\ref{resc}) reduces to the  $\nu_{16}$ 
resonance condition where we regard the planet as an asteroid that interacts with Jupiter and Saturn,
instead of the disk and companion star. $\omega_{\rm ds}$ is regarded
as the sum of precession rates due to the mutual torques involving the dominant components, Jupiter and Saturn.  Since the angular momentum of
 an asteroid is very small compared the angular momentum of Jupiter or Saturn (implying in effect $J_{\rm p} \ll J_{\rm d}$), we see
 from Equations (\ref{imax}) and (\ref{dm}) that the inclination change is very large and is confined
 to a narrow range of perturbing mass at fixed position that translates into a narrow range of
 radii at fixed mass. 
 
 The case of a disk-planet system is somewhat different, since the giant planet mass is not extremely
 small compared to the disk mass. For the upper curve plotted in left panel of Fig.~\ref{deltaimass}, near the peak
 value of $A_{\rm pd}/i_0$, where $M_{\rm d} \simeq 2\times10^{-3}M$, 
 the angular momentum ratio is $J_{\rm p}/J_{\rm d} \simeq 0.4$.
 As a result, the value of $A_{\rm max}/i_0$ is not
 very large and the dimensionless 
 resonance width in terms of disk mass $\delta M_{\rm d}/M_{\rm d}$ is not small.
 Consequently, the effects of the resonance extend over a broad range of disk mass.

 We define a dimensionless measure of
 the closeness to the resonance condition (\ref{resc}) (the detuning parameter) by
  \begin{equation}
 D(M_{\rm d})=    \frac{\omega_{\rm pd} \left(1+ J_{\rm p}/J_{\rm d}  \right)+\omega_{\rm p s}}{\omega_{\rm d s}}-1,
 \label{DM}
 \end{equation}
 where $\omega_{\rm pd}$ and $J_{\rm d}$ are linear functions of $M_{\rm d}$.
 For  $D(M_{\rm d})$ positive (negative), the relative nodal phasing of the planet and disk is librating (circulating).
 For $D(M_{\rm d})$  negative, the planet orbit inclination relative to the disk is of order $i_0$, as a 
 consequence of the circulation. For $D(M_{\rm d})$ large and positive, the planet inclination relative to the disk is much less than $i_0$
 and alignment occurs. For $|D(M_{\rm d})|$ of order unity,  the  planet inclination relative to the disk is of order $i_0$, as a consequence
 of the secular resonance.
 
\begin{figure}
\begin{center}
\includegraphics[width=8.0cm]{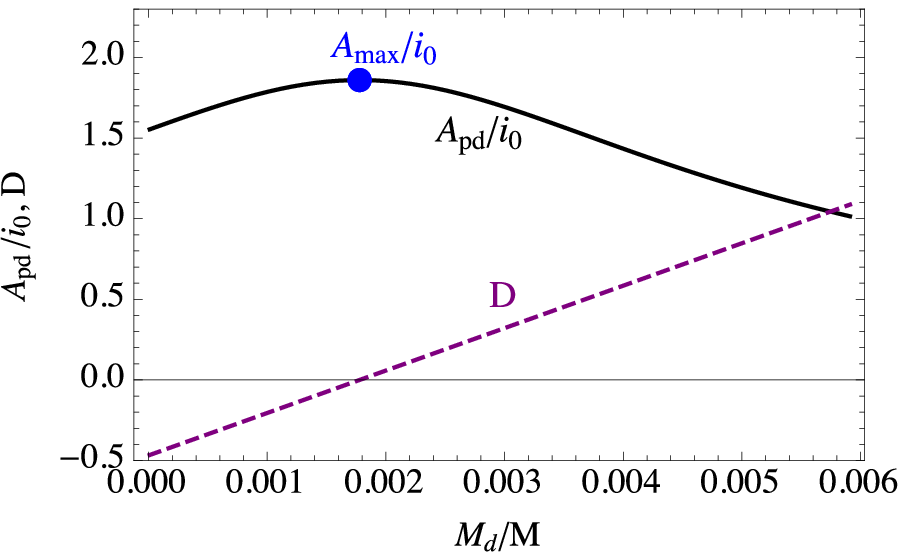}
\end{center}
\caption{The  solid line  plots the tilt oscillation amplitude of the planet orbit relative to the disk, $A_{\rm pd}/i_0$, as a function of disk mass (normalized by binary mass $M$) for the  case of the dashed and overlapping solid line
 in the left panel of Fig.~\ref{deltaimass}. 
 The dot plots the predicted maximum
 value of $A_{\rm pd}/i_0$ according to Equation (\ref{imax}).
 The dashed line plots the secular resonance
 detuning parameter $D(M_{\rm d})$ of Equation (\ref{DM}) that vanishes for the disk mass at maximum $A_{\rm pd}/i_0$. Planet--disk alignment occurs for large $D(M_{\rm d})$.
  }
\label{fig:rescon}
\end{figure}

In Fig.~\ref{fig:rescon}, we plot as a solid line the case of the dashed and overlapping solid line
 in the left panel of Fig.~\ref{deltaimass}.  The dashed line plots the  resonance
 condition (\ref{DM}) and the dotted line plots the predicted maximum value of
 $A_{\rm max}/i_0$ in Equation (\ref{imax}) with $J_{\rm d}$ evaluated at the resonance
 disk mass (although the plot extends over different disk masses). 
 As expected, the peak misalignment occurs for a disk mass where $D(M_{\rm d})=0$
 and therefore where Equation (\ref{resc}) is satisfied. The peak value agrees with the predicted value.
 In this case,  the resonance condition is satisfied for a disk mass
 that is about twice  the planet mass.

 The nonmontonic behavior of $A_{\rm pd}/i_0$ as a function of disk mass
 is then a consequence of the secular resonance. That is, the increase in relative
 planet-disk tilt with increasing disk mass is due to the effects of the secular
 resonance driven by the disk and binary companion that causes misalignment as measured by $A_{\rm pd}/i_0$ to have order unity values ($\sim 0.5$ or larger)  in the cases we have considered.  
The results of this section suggest that maintaining coplanarity by means of gravitational torques
between the planet and the disk appears to only be possible if the mass of the
disk is very large compared to that of the planet, or if the disk
extends very close to the planet. Smaller mass planets open a
  smaller gap in the disk and are more likely
  to remain coplanar.

\section{Two Planets in a Binary System}
\label{tp}

In this section we consider a binary star system with two misaligned
planets around one star. We are essentially replacing the disk in
Section \ref{secular} with an outer planet.  By doing so, we can
easily carry out simulations that can be compared with the secular
theory of Section \ref{secular} to test its accuracy. Of course, a
disk is an extended body that can lie closer to both the binary
companion and the inner planet than can a point mass object such as a
planet.  This test then involves a somewhat different situation than
having a disk. But it can give a general estimate of the accuracy of
the secular theory for the range of parameters involved in the planet--disk case.

\subsection{Secular Interactions}
\label{sec}

We adopt an equal mass binary system with $M_1=M_2=0.5\,M$ and planets ${\rm p1}$ and ${\rm p2}$
with masses $M_{\rm p1}=M_{\rm p2}=1 \times 10^{-3}M$ orbiting at radii $a_{\rm
  p1}=0.1\,a$ and $a_{\rm p2}=0.2\,a$, respectively. 
We apply the methods described in Section~\ref{secular}, but replace the disk with a
point mass with mass $M_{\rm p2}$ at radius $a_{\rm p2}$ and relabel the planet ${\rm p}$ as planet ${\rm p1}$. Thus, we
replace the angular momentum of the disk component with
\begin{equation}
J_2=M_{\rm p2}a_{\rm p2}^2\Omega(a_{\rm p2}),
\end{equation}
replace the coupling coefficient $C_{\rm pd}$  by
\begin{equation}
C_{\rm p1 p2}=GM_{\rm p1}M_{\rm p2}K(a_{\rm p1},a_{\rm p2}),
\end{equation}
and replace $C_{\rm ds}$  by
\begin{equation}
C_{\rm p2 s}=GM_{\rm p2}M_2K(a_{\rm p2},a).
\end{equation}

The planet orbits are initially coplanar
  but misaligned with respect to the binary orbital plane. We consider three different
initial binary misalignments of $10^\circ$, $20^\circ$, and
$30^\circ$.  
Because the secular model equations are linear, each is just a
scaled version of the others.  Fig.~\ref{twoplanet} shows the tilt and
phase angle evolution for each planet. The orange line corresponds to
the inner planet ${\rm p1}$ and the red line the outer planet  ${\rm p2}$.

\subsection{4--body simulations}
\label{nbody}

\begin{figure*}
\begin{center}
\includegraphics[width=5.9cm]{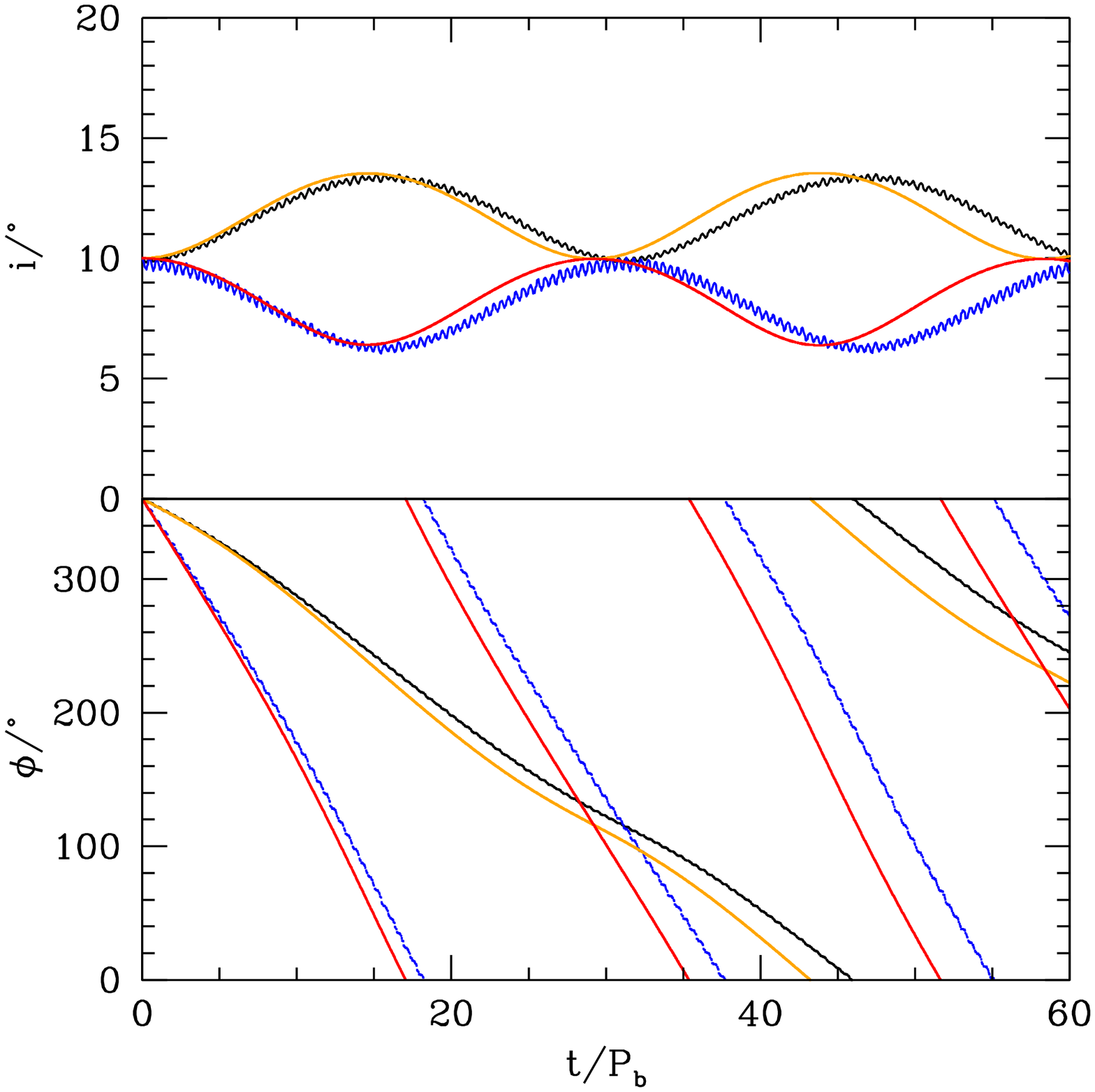}
\includegraphics[width=5.9cm]{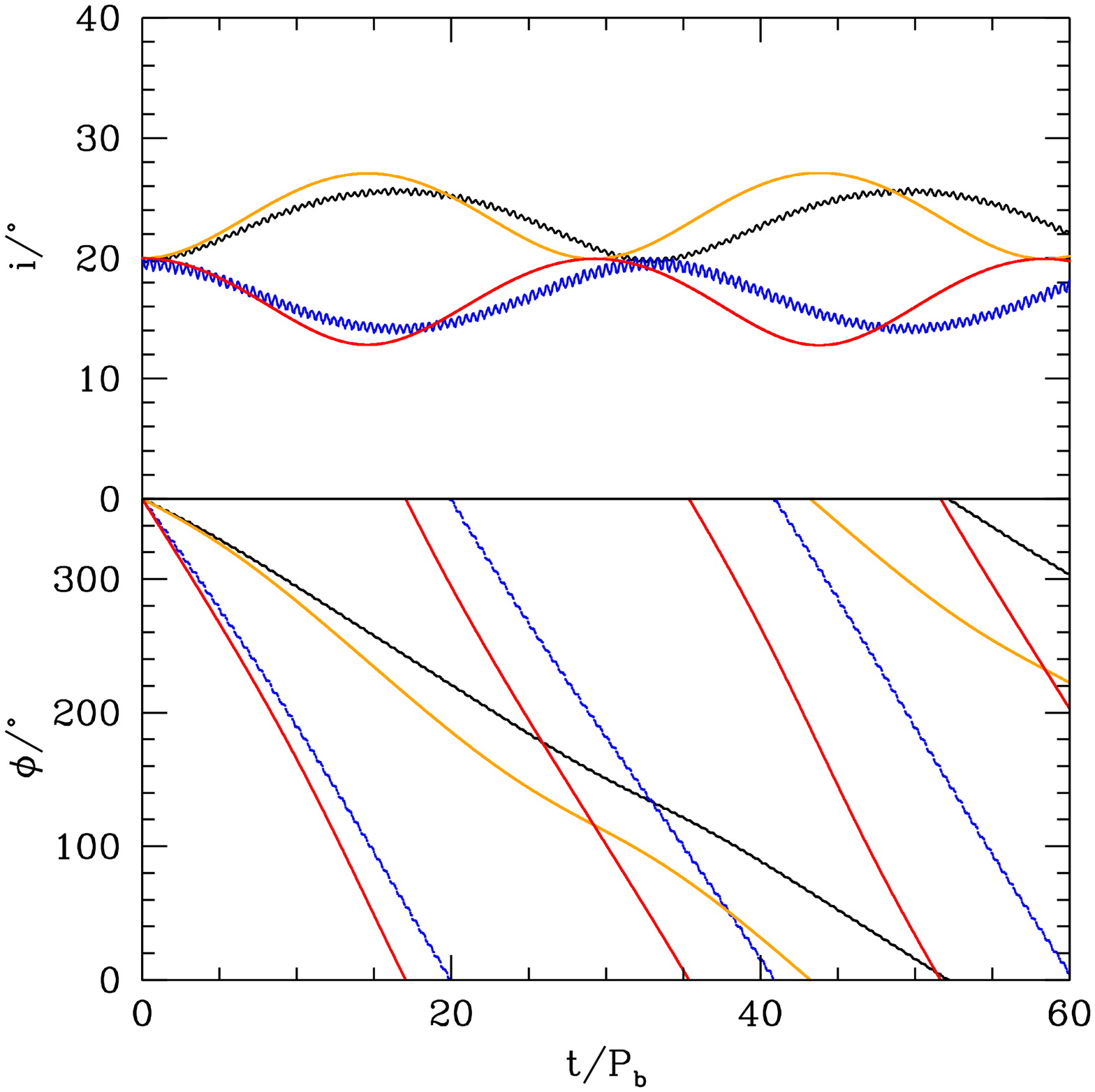}
\includegraphics[width=5.9cm]{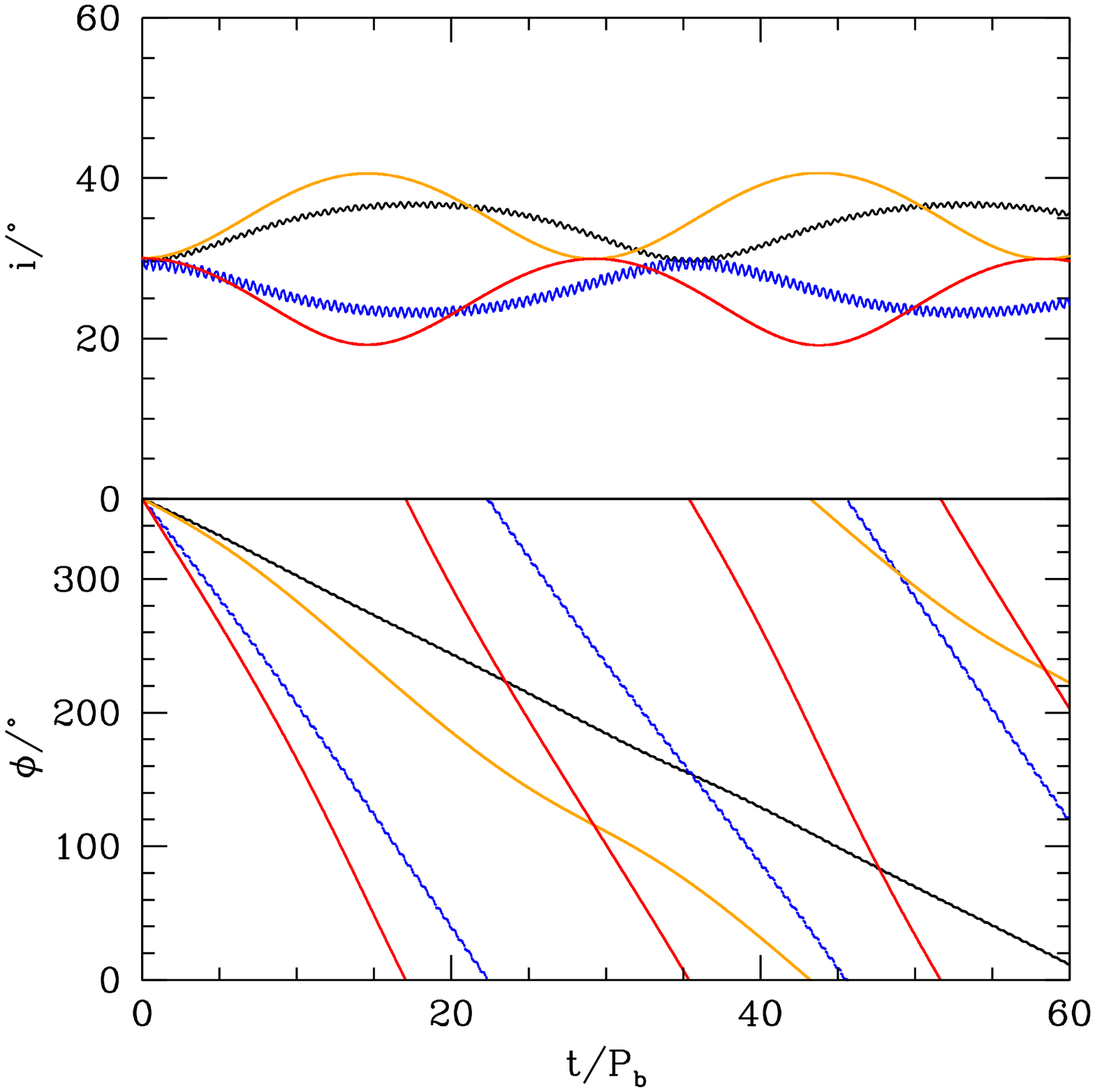}
\end{center}
\caption{Tilt (upper panels) and phase angle (lower panels) evolution
  of a two planet system around one component of an equal mass binary. The initial
  inclination of both planets is $i_0=10^\circ$ (left), $20^\circ$
  (middle) and $30^\circ$ (right). The orange and red  lines show
  the inner and outer planets respectively for the secular model
  described in Section~\ref{sec}. The black and blue lines show the
  inner and outer planets respectively for a simulation
  described in Section~\ref{nbody}.  The planets both have mass
  $1 \times 10^{-3}\,M$ and begin at radii $a_{\rm p1}=0.1\,a$ and $a_{\rm p2}=0.2\,a$.}
\label{twoplanet}
\end{figure*}

We describe the numerical simulations where we integrate the
gravitational force equations  in time. We work in the inertial frame and fix
the orbits of the stars.  As in the
planet--disk secular model, we do not allow the orbit of the binary to evolve
under the influence of the other objects.
We apply a Cartesian coordinate system
$(x,y,z)$ with the origin at the binary center of mass, the $x-$axis
along the line joining the initial positions of the two stars, and the
$z-$axis along the binary rotation axis.   Star 1 that is central to the two planets has position
\begin{equation}
{\bf r}_1(t)=a_1(\cos t, \sin t,0),
\end{equation}
where $a_1=a M_2/M$ is its binary orbital radius. The companion star 2 has position
\begin{equation}
{\bf r}_2(t) =-a_2(\cos t, \sin t,0),
\end{equation}
where $a_2=a M_1/M$ is its orbital radius. For an equal mass binary, we have that
$a_1=a_2=a/2$.  The position of the inner planet is denoted by ${\bf r}_{\rm p1}(t)$ and the
outer planet is denoted by ${\bf r}_{\rm p2}(t)$.  We solve the differential equations for
the position of the inner planet
\begin{equation}
\frac{d^2 {\bf r}_{\rm p1}}{d t^2}
+ \frac{G M_1( {\bf r}_{\rm p1} -  {\bf r}_1  ) } {  |{\bf r}_{\rm p1}-{\bf r}_1|^{3}}
+ \frac{G M_2( {\bf r}_{\rm p1} -  {\bf r}_2  ) } {  |{\bf r}_{\rm p1}-{\bf r}_2|^{3}}
+ \frac{G M_{\rm p2}( {\bf r}_{\rm p1} -  {\bf r_{\rm p2}}  ) } {  |{\bf r}_{\rm p1}-{\bf r_{\rm p2}}|^{3}}=0
\end{equation}
and for the outer planet
\begin{equation}
\frac{d^2 {\bf r}_{\rm p2}}{dt^2}
+ \frac{G M_1( {\bf r}_{\rm p2} -  {\bf r}_1  ) } {  |{\bf r}_{\rm p2}-{\bf r}_1|^{3}}
+ \frac{G M_2( {\bf r}_{\rm p2} -  {\bf r}_2  ) } {  |{\bf r}_{\rm p2}-{\bf r}_2|^{3}}
+ \frac{G M_{\rm p1}( {\bf r}_{\rm p2} -  {\bf r_{\rm p1}}  ) } {  |{\bf r}_{\rm p2}-{\bf r_{\rm p1}}|^{3}}=0.
\end{equation}
Planets ${\rm p1}$ and  ${\rm p2}$  begin on the $x$--axis at radii $a_{\rm p1}=0.1\,a$ and $a_{\rm p2}=0.2\,a$.
from star 1. The planets are given an initial velocity in the $y$--$z$ plane
such that they are on a circular Keplerian but tilted orbits about the
central star 1. The resulting evolution of the inclination of the two planets is shown
in the black and blue  lines in Fig.~\ref{twoplanet}. 

The numerical simulations and the secular evolution predicted in the
previous section show very similar behavior for low inclination angles
in Fig.~\ref{twoplanet}. Thus, we conclude from this section that the
secular evolution model in Section~\ref{secular} provides an
approximately accurate description of tilt evolution for small tilts
$i \lesssim 20^{\circ}$.

\section{Hydrodynamical Simulations}
\label{sec:sph}

\begin{figure}
\begin{center}
\includegraphics[width=8.0cm]{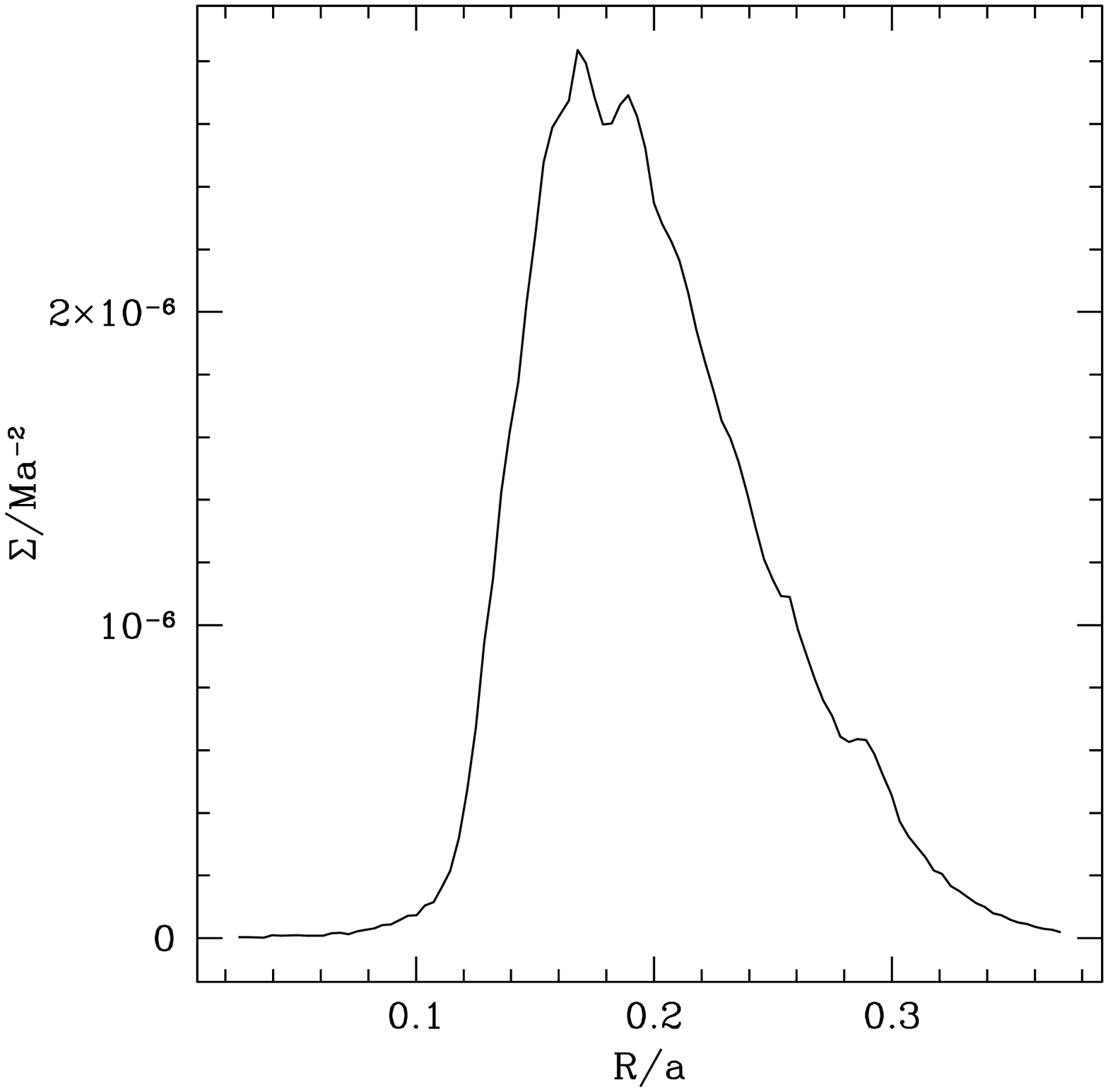}
\end{center}
\caption{Surface density profile at a time of 10 binary orbits for a
  coplanar disk with an initial mass of $10^{-6}M$. The surface density $\Sigma$
  is normalized by $M/a^2$, the binary mass divided by the square of the binary separation.
  The simulations begin with a planet embedded in a disk without a gap, as discussed in Section \ref{sec:sph}. }
\label{sigma}
\end{figure}

\begin{figure*}
\begin{center}
\includegraphics[width=8.0cm]{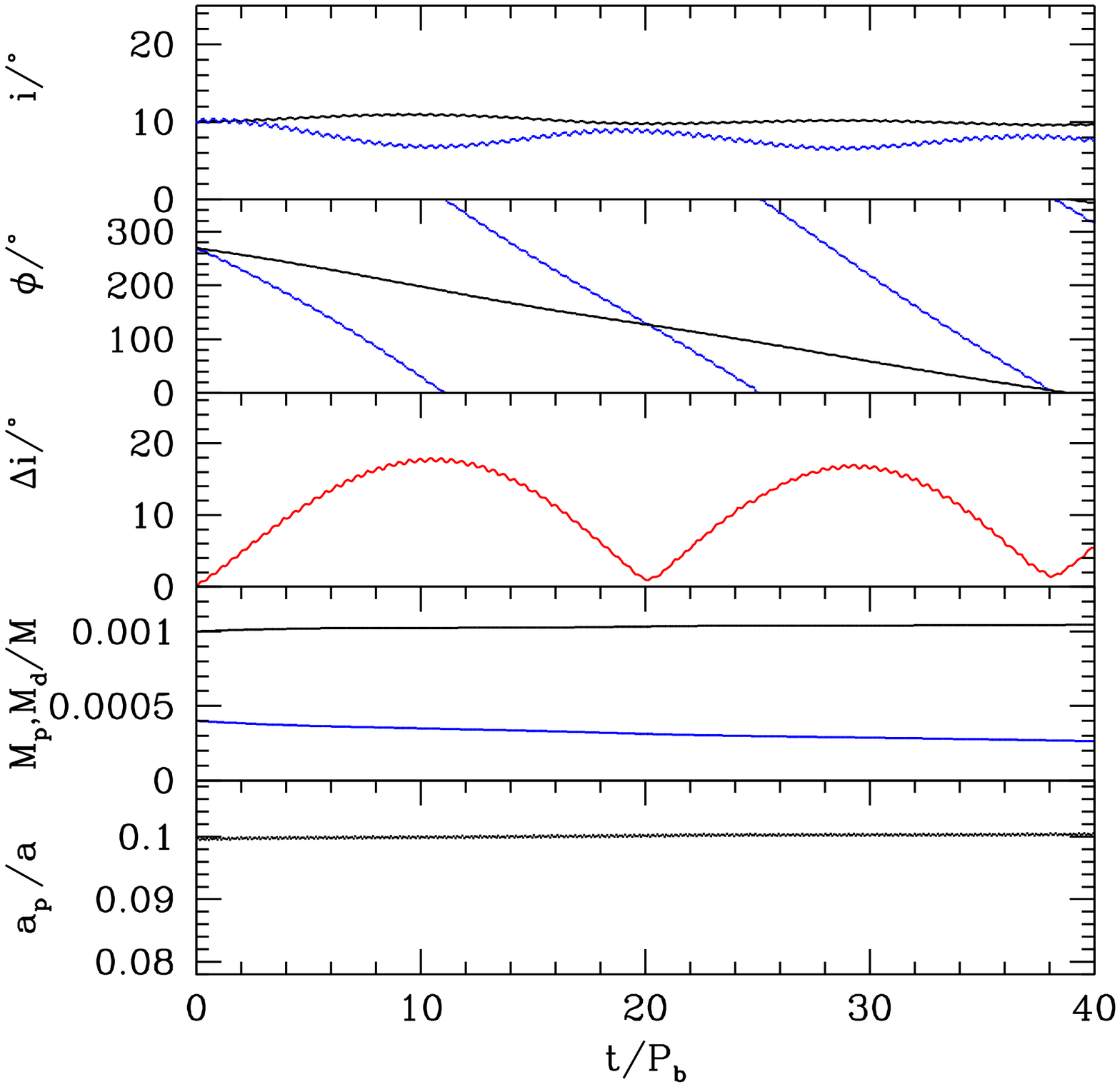}
\includegraphics[width=8.0cm]{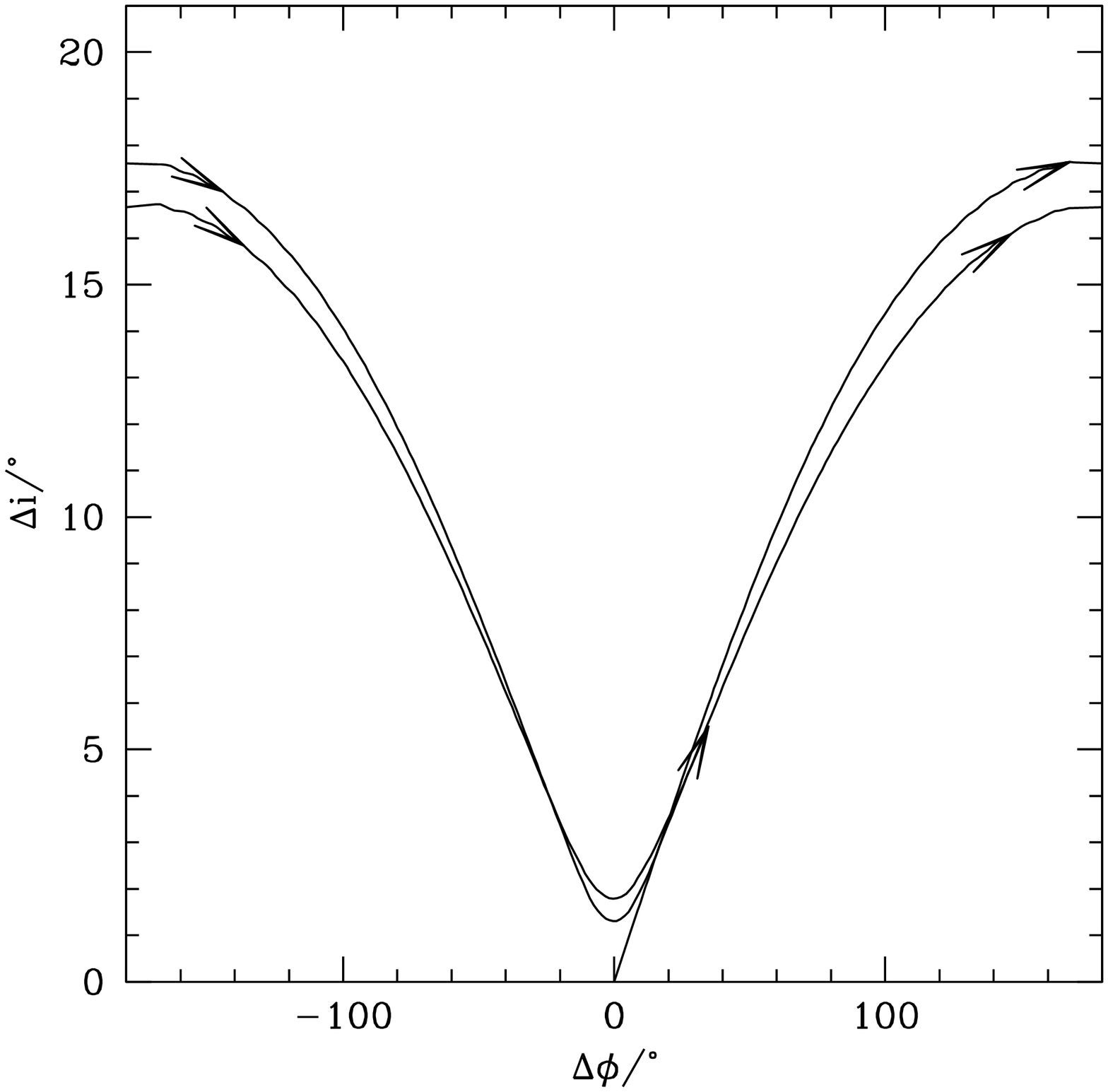}
\end{center}
\caption{SPH simulations of a disk with a planet in a misaligned
  binary system for a system with an initial inclination of
  $i=10^\circ$.  The plotted time $t$ is the time after an initial disk adjustment
  phase of $10 P_{\rm b}$ that begins with a planet embedded in a disk without a gap.
  The disk mass at $t=0$ is $4 \times 10^{-3}\, M$ and the planet
  mass is $1 \times 10^{-3}\,M$, where $M$ is the mass of the binary. 
  Left: Evolution of the planet (black) and outer
  disk at a radius of $0.2\,a$ (blue). The top graph plots as a function of time the
  inclination, the second graph plots the phase angle, and the third graph
  plots the relative planet--disk tilt, $\Delta i= |W_{\rm p}-W_{\rm
    d}|$ evaluated at $R=0.2a$. The fourth graph plots the mass of the
  planet and the disk and the bottom graph plots the the semi--major
  axis of the planet. Right: Phase portrait of the relative
  planet--disk tilt and phase angle averaged over the previous binary
  orbit versus nodal phase difference, $\Delta \phi=\phi_{\rm p} -
  \phi_{\rm d}$.  }
\label{sph}
\end{figure*}

In this Section we describe 3D hydrodynamical simulations to model 
planet--disk--binary systems.  We use the smoothed particle
hydrodynamics (SPH; e.g., \citealt{Price2012a}) code {\sc phantom}
\citep{PF2010,LP2010,Nixon2012,Nixonetal2012b,Nixonetal2013}.  The
binary star system, disk, and planet parameters are summarized in
Table~\ref{tab}.  We consider an equal mass binary star system, with
total mass $M=M_1+M_2$ that has a circular orbit in the $x$-$y$ plane
with separation $a$. The mass of the planet is $M_{\rm p}=1 \times 10^{-3}\,M$
and its initial distance from the central star is $0.1\,a$. We
choose the accretion radius about each star to be $0.025\,a$ and about
the planet to be $0.005\,a$.  We have found that the simulation results
do not vary significantly for smaller values of these accretion radii. With these
values we are able to run the simulations faster than with smaller values.
Particles that fall within this radius
are removed from the simulation and their mass and momentum are added
to the accreting object. 

 The disk is locally isothermal with sound speed $c_{\rm s}
\propto R^{-3/4}$ and 
$H/R=0.036$ at $R=R_{\rm in}$. With these
parameters, $\alpha$ and $\left<h\right>/H$ are constant over the disk
\citep{LP2007}. The \cite{SS1973} $\alpha$ parameter is taken to be
0.05. We implement the disk viscosity by adapting the SPH artificial
viscosity according to the procedure described in \cite{LP2010}, using
$\alpha_{\rm AV} = 0.76$ and $\beta_{\rm AV} = 2.0$. The disk is
resolved with shell--averaged smoothing length per scale height
$\left<h\right> /H \approx 0.66$.

In order to simulate a planet--disk system, we must choose
an appropriate initial disk density distribution. In the 
case of coplanar systems, we can choose a disk profile
with or without an initial gap. As the system evolves,
a reasonably stable equilibrium gap structure is produced typically
in less than 100 planet orbits \cite[e.g.,][]{Bateetal2003}. 
In case of a planet--disk system that is misaligned with respect to a binary, the initial conditions
need to be more accurately established because the
disk could precess substantially 
before an equilibrium gap is established. That is, an initially aligned
planet--disk system may  not retain this desired initial
relative alignment until the equilibrium gap is established.
A further problem is that if we start without a  gap, the planet can
gain substantial mass before gap opening and thereby prevent us from
easily controlling the  planet mass when an equilibrium gap is established.

To mitigate these problems, we have developed a procedure
in which we first simulate a coplanar planet--disk--binary system
with a  very  low mass initial disk, $M_{\rm d}= 1\times 10^{-6}M$. The disk starts without a gap.
The disk initially extends from $R_{\rm
  in}=0.025\,a$ to $R_{\rm out}=0.25\,a$.  
 The outer disk radius
$R_{\rm out}$  is close to the tidal truncation radius of a disk in
a coplanar binary system \citep[see][]{Paczynski1977}.
The initial surface density follows the power law $\Sigma \propto
R^{-3/2}$. We evolve this coplanar simulation for 10 binary orbital periods.
By this time, a stable disk gap structure is created while the planet
mass and orbit remain nearly unchanged. Fig.~\ref{sigma} shows the resulting
surface density profile.

It can be shown that  for a locally isothermal, non-selfgravitating disk (as we assume), and fixed orbits of the binary and planet,
the disk density profile shape is independent of disk mass. 
We then rescale the density distribution obtained after 10 binary orbital periods to achieve the desired  disk mass.
We also tilt both the planet and disk relative to the binary orbital plane by the desired
angle $i_0=10^\circ$ so that they are mutually coplanar.  
This planet--disk configuration then serves as the initial conditions for the subsequent simulation of a planet--disk system that is
misaligned with respect to the binary orbital plane.

After the initial coplanar simulation completes in 10 binary orbital periods, much
of the disk that lies interior to the orbit of the planet has been accreted onto the central star.
The resulting disk lies primarily exterior to the planet orbit.
Due to the approximations made in the establishing the initial conditions for the tilted system, 
the  system makes some initial readjustment to the tilt and increased 
disk mass.
The tidal forces are weaker for a tilted disk than a coplanar one  \citep{Lubowetal2015}.
Consequently, the disk
expands outwards somewhat. However, since the tilt is small, this should
not be a large readjustment.  Another effect is that the increase in the disk mass results in outward gravitational forces on the planet
 that cause its orbit to initially expand slightly.

We examine disks with three different masses, $M_{\rm
  d}=4 \times 10^{-4} M$, $4 \times 10^{-3}M$ and $6\times10^{-3}M$, for the initially tilted disk.  The simulations start
with $5 \times 10^{5}$ SPH particles in the initial coplanar (pre-tilted) configuration.

\begin{table*}
\caption{Parameters of the initial disk conditions and for a circular equal
  mass binary with total mass, $M$, and separation, $a$.} \centering
\begin{tabular}{lllll}
\hline
Simulation Parameter & Symbol & Values\\
\hline
\hline
Mass of binary component &  $M_1/M = M_2/M$ & 0.5  \\
Accretion radius of the binary masses & $R_{\rm acc}/a$    & 0.025   \\
\hline
Initial disk mass & $M_{\rm di}/M$ & [$4 \times 10^{-4}, 4 \times 10^{-3}, 6 \times 10^{-3}$ ]\\
Disk viscosity parameter & $\alpha$ & $0.05$ \\
Disk aspect ratio & $H/R (R=R_{\rm in})$ & 0.036 \\
   & $H/R (R=R_{\rm out})$ & 0.02 \\
Initial disk inclination & $i/^\circ$ & $10$  \\ 
\hline
Planet Mass & $M_{\rm p}/M$ &  $1 \times 10^{-3}$ \\
Initial planet inclination & $i_{\rm p}/^\circ$ &10  \\
Initial planet separation to primary & $a_{\rm p}/a$ & 0.1  \\
Accretion radius of the planet & $R_{\rm p,acc}/a$ & 0.005 \\
\hline
\end{tabular}
\label{tab}
\end{table*}

We first consider the evolution of a small mass disk with a total mass of
$4 \times 10^{-4} M$.
The left panel of Fig.~\ref{sph} shows the evolution of the
system in time.  The upper two graphs show the inclination and phase
angle evolution for the disk and the planet.
The disk remains largely unwarped (flat) during the evolution. We plot
the disk evolution at a representative radius of $0.2\,a$. 
The black lines show the evolution of the planet and the blue
lines the disk. 
After the first oscillation, $\Delta
i$ does not return to zero
but
remains approximately periodic.
Unlike the oscillations in the secular model, the
oscillations in the SPH simulations are damped through the disk
viscosity. 

In comparing these results to the left panel of
Fig.~\ref{planetdisc}, we see that the periods of the tilt oscillations
in the SPH simulation and secular model have similar values of about $20 P_{\rm b}$. 
There is some difference in the tilt amplitudes $i(t)$. They  are likely due to differences
in our parameter choices for the secular model, such as the disk outer radius. 
 The red line in the middle graph plots the relative
inclination between the planet and the disk, $\Delta i= |W_{\rm
  p}-W_{\rm d}|$. 
  The relative inclination of the planet to the disk $\Delta i$ reaches values
of about 1.8 times  $i_0$  or $18^{\circ}$ that is similar to the value of about $17^{\circ}$
as implied by in Fig.~\ref{phasepor} for this disk mass.
The fourth graph shows the mass of each component
and the bottom graph shows the orbital semi--major axis of the
planet. There is little accretion or migration of the planet caused by this
small mass disk.

The right panel of Fig.~\ref{sph} plots the relative disk--planet
tilt  as a function of nodal phase
difference $\Delta \phi=\phi_{\rm p} - \phi_{\rm d}$ with the individual phases plotted in
the second graph of the left panel.
The system
starts at $\Delta \phi=0$ and $\Delta i=0.$ The phase wraps from
$+180^\circ$ to $-180^\circ$, as the system evolves.  The system is
circulating, rather than librating.
In both the secular model and the SPH simulation,
the relative phasing between the planet and disk is fully circulating (see Fig.~\ref{phasepor} and right panel of Fig~\ref{sph}).

\begin{figure*}
\begin{center}
\includegraphics[width=8.0cm]{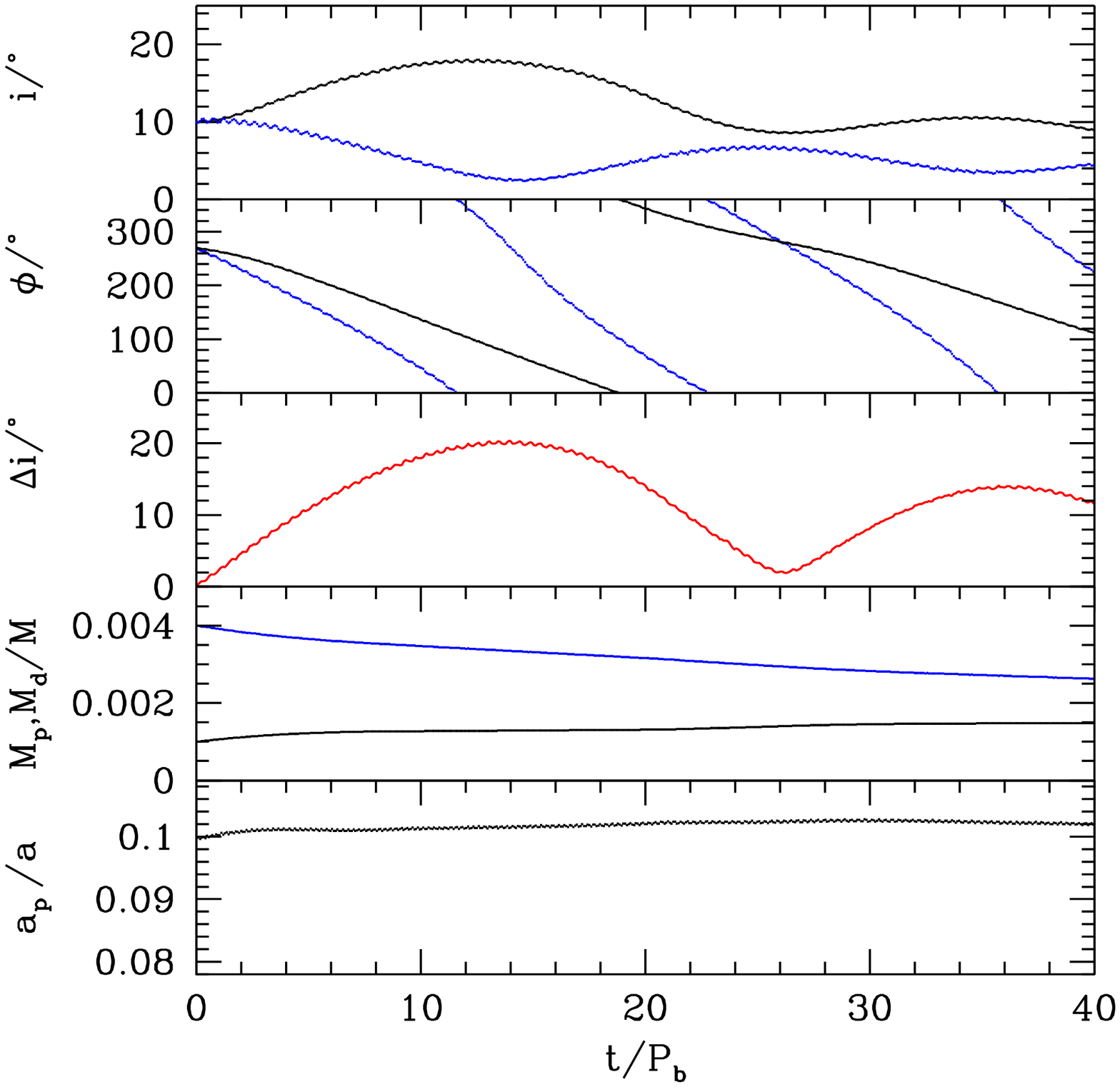}
\includegraphics[width=8.0cm]{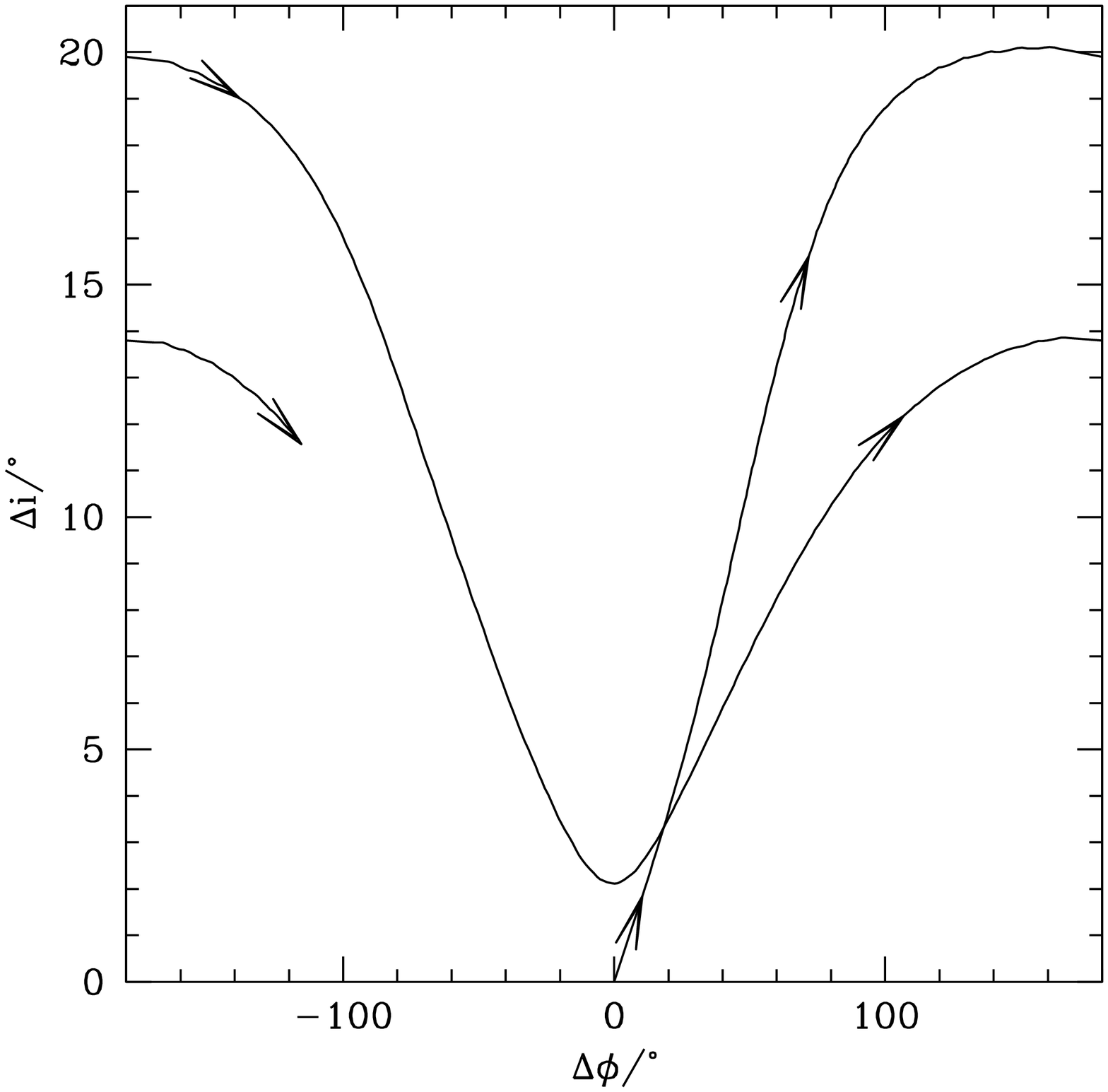}
\end{center}
\caption{Same  as Fig.~\ref{sph}, but for a disk mass of
  $4 \times 10^{-3}M$ at $t=0$, where $M$ is the mass of the binary. }
\label{sph4}
\end{figure*}

Fig.~\ref{sph4} shows the evolution of planet--disk system with a
higher initial disk mass of $4 \times 10^{-3}M$. The tilt oscillations
are larger with the higher disk mass, as was seen for the secular model in Fig.~\ref{planetdisc}.  
There is more
migration of the planet with the higher mass disk, but the amount of migration over the course of the simulation is small. 
The planet gains about 50\% more mass by the end of the simulation which is about 1/3 of the amount of gas
that has been lost from the disk.
Gas giant planets in this general mass range in coplanar planet--disk systems typically accrete most of the gas that flows past their orbits
\citep{Lubow06}.
As seen in Fig.~\ref{planetdisc} and Fig.~\ref{sph4}, the secular model and the initial oscillation in the simulations 
show a similar peak value of $i_{\rm p} \sim 2 i_0$. On the other hand, the inclination of the disk
reaches lower values than predicted by the secular model. 

In the right panel of Fig.~\ref{planetdisc} , the secular  model predicts libration at this disk mass, rather than
the circulation found in the SPH simulations. It is clear from  Fig.~\ref{sph4} that the phasing
of the planet and disk is not locked, even at early times. 
This comparison indicates that peak planet--disk misalignment, as seen in Fig.~\ref{fig:rescon},
occurs at a higher value of the disk mass than the secular model suggests, higher than $M_{\rm d}=4 \times 10^{-3} M$.
For $t  \la 20 P_{\rm b}$, the planet--disk misalignment at higher disk masses is even stronger than estimated by the secular
model with the adopted standard parameters.
This shift could be a consequence, for example, of underestimating the disk outer radius as $0.25 a$ for the secular model compared with possibly larger values, as suggested by Fig.~\ref{sigma}.
The SPH results indicate that the transition from circulation to libration occurs at a disk mass of about $5 \times 10^{-3}M$.  Such a shift would in effect change the plot of $D(M_{\rm d})$ in Fig.~\ref{fig:rescon} from passing through
zero at  $M_{\rm d}  \simeq 0.002 M$ to passing through
zero at $M_{\rm d} \simeq 0.005 M$, about 5 times the planet mass, where $A_{\rm pd}(M_{\rm d})/i_0$  would be maximum.

At later times, additional differences from the secular model are
likely due to accretion of gas by planet  as well as the disk tilt decay by dissipation in the simulations
that are not included in the secular model. The disk viscosity can also result in the decay of the planet--disk
relative tilt \citep{Lubow2001}.
The planet gains both mass and momentum from the disk as it accretes gas.
Consequently, its orbit tilt becomes more aligned with the disk that is becoming more
coplanar with the binary at later times. However,  over the time span  of the simulations, the tilt of the planet orbit
relative to the disk is generally of order the initial tilt $i_0$. 

\begin{figure*}
\begin{center}
\includegraphics[width=8.0cm]{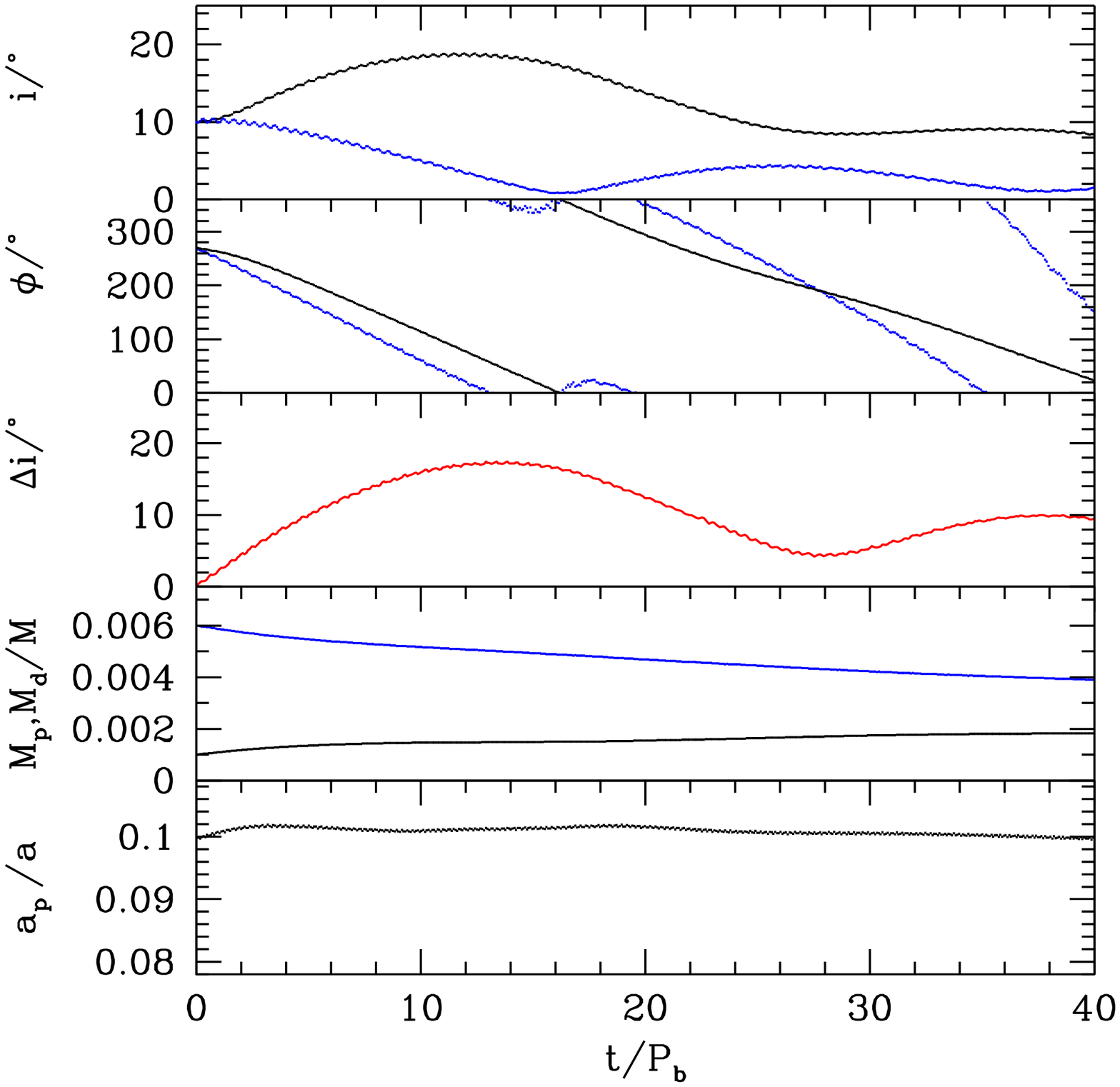}
\includegraphics[width=8.0cm]{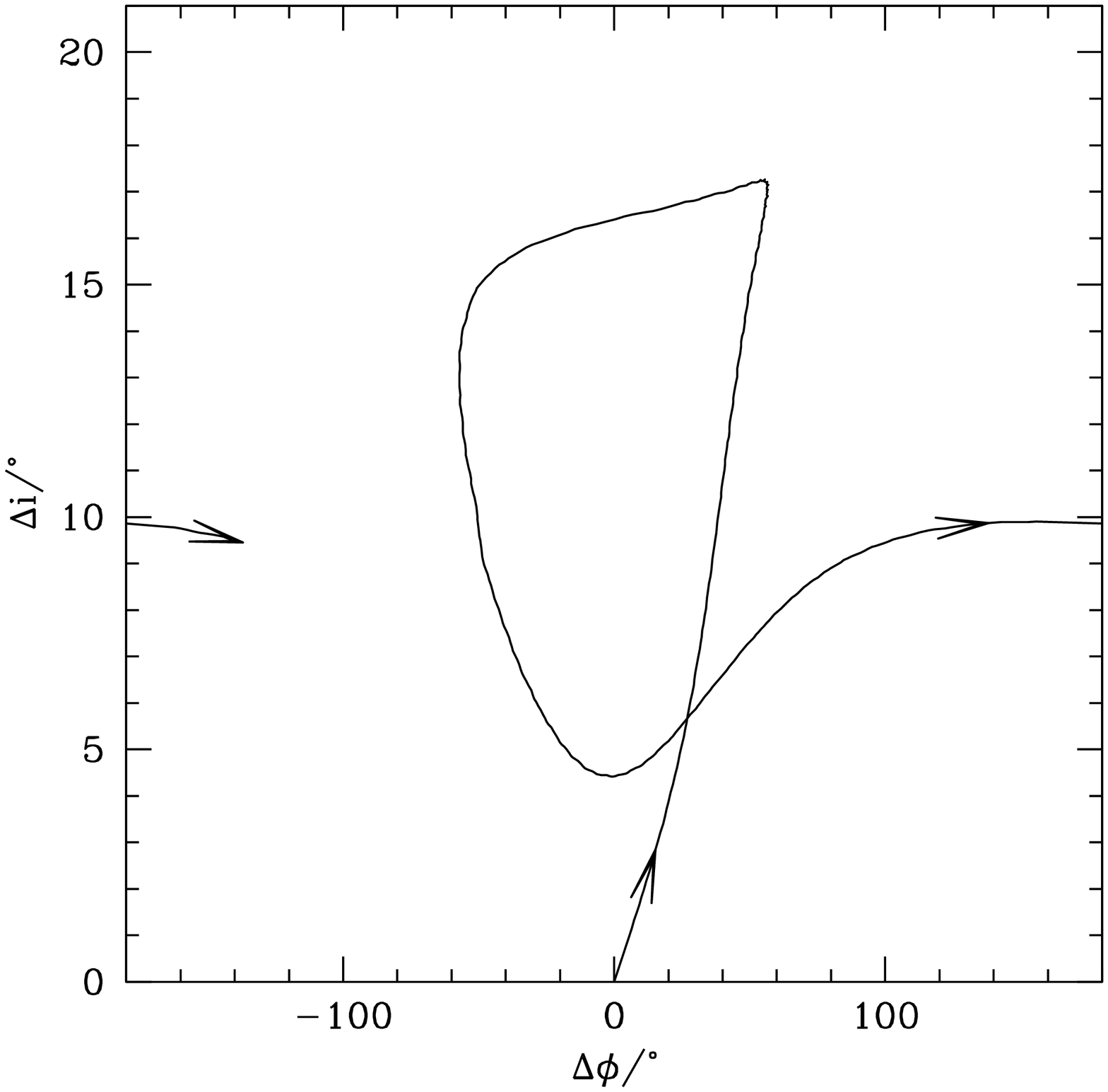}
\end{center}
\caption{Same as Fig.~\ref{sph}, but for a disk mass of
  $6 \times 10^{-3}M$ at $t=0$, where $M$ is the mass of the binary.   As seen in the right panel, the phase portrait
  shows a transition from libration to circulation as the disk mass decreases.
  }
\label{sph2}
\end{figure*}

\begin{figure}
\begin{center}
\includegraphics[width=8.0cm]{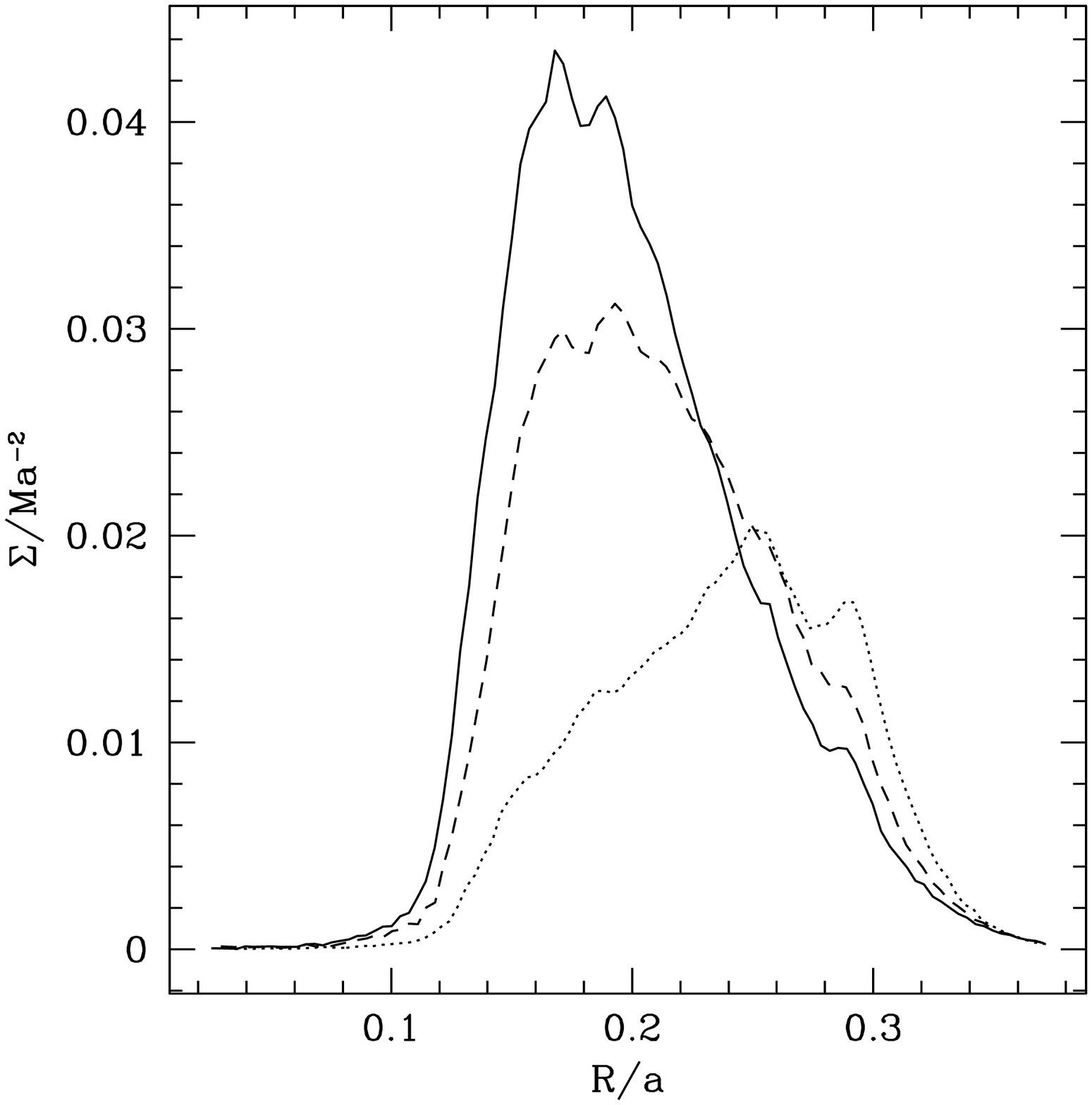}
\end{center}
\caption{Disk surface density profiles at 
times $t$ after an initial disk adjustment phase of $10  P_{\rm b}$. The densities are plotted for $t=$ 0 (solid line), $6 P_{\rm b}$ (dashed line), and $40 P_{\rm b}$ (dotted) for the model plotted in Fig.~\ref{sph2}. 
 The surface density $\Sigma$
  is normalized by $M/a^2$, the binary mass divided by the square of the binary separation.
  Note that in the adjustment phase, the simulations begin with a planet embedded in a disk without a gap, as discussed in Section \ref{sec:sph}. The planet is located at $R=0.1 a$ at $t=0$. The disk has a mass of $6 \times 10^{-3} M$
  at $t=0$ where $M$ is the binary mass.
   }
\label{sigmaev}
\end{figure}

Fig.~\ref{sph2} shows results for a somewhat higher disk mass of $6 \times 10^{-3}M$. 
As a consequence of this higher disk mass, the planet orbital radius
initially expands more than in the lower disk mass cases, the planet gains about an additional $\sim 80\%$ of its initial mass by
the end of the simulation.
The disk tilt decays to a nearly coplanar configuration with the binary orbital plane.
Again, the planet gains both mass and momentum from the disk as it accretes gas.
Consequently, its orbit tilt becomes more aligned with the disk that is nearly
coplanar with the binary at later times. 
 
The right panel of Fig.~\ref{sph2} plots the relative
disk--planet tilt $\Delta i= |W_{\rm p}-W_{\rm d}|$ as a function of
nodal phase difference $\Delta \phi=\phi_{\rm p} - \phi_{\rm d}$.  The
system starts at $\Delta \phi=0$ and $\Delta i=0.$ After the initial
time, $\Delta i$ does not return to zero and is not very periodic.
The system is initially librating and then transitions to circulating
at late stages of the evolution when the disk has lost most of its initial
mass.  Essentially, the system is passing through the
cycles plotted in Fig.~\ref{phasepor} with decreasing disk mass over time,   starting at
a loop for high disk mass, reaching the near flat-top
at the edge of libration, and breaking to circulation at
low disk mass.

The disk surface density evolution is plotted 
in Fig.~\ref{sigmaev}. The solid line plots the
initial surface density distribution that is obtained by rescaling the density distribution obtained
from the evolved coplanar configuration (plotted
in Fig.~\ref{sigma}) so that the initial disk mass is $6 \times 10^{-3}M$. The density distributions at later times show some further outward expansion along with a decrease in disk mass. The plots show that the deep disk gap about the planet that is located near $r=0.1 a$ is maintained over time. 

\section{Discussion }
 \label{disc}

We have found in Section \ref{sec:secres} that a secular resonance
 plays 
an important role in the misalignment process.
Secular resonances have been previously considered in the context of disk--planet systems.
Ward (1981) showed 
that secular resonances caused by planets could have swept across portions of the early solar system.
The sweeping is due the gravitational effects of the gaseous disk, even if the planets do not migrate. 
The reason for the sweeping is that even a minimum-mass solar nebula can have an important influence on the relevant precession rates. The role of the disk is to modify the precession rates.
As the nebula disperses, the precession rates vary, along with the resonance locations. 
Solid bodies, such as the terrestrial planets, can be driven into significantly eccentric and inclined orbits
as these resonances pass through their orbital locations. 

In another study,
\cite{Lubow2001} examined the response of a gaseous disk to secular driving
by two planets that are misaligned with respect to a disk. Because the precessional forcing period is long
compared to the sound crossing time, the disk response is global in the form
of a large scale warp.
In these previous models the precessional modes that drive the resonances 
are due to compact objects (planets). In the secular resonance described in Section \ref{sec:secres},
the disk and binary companion star drive the secular resonance.

Also, unlike the case examined  by Ward (1981) the angular momentum of the smallest mass
object in the system, the planet in this paper, is not extremely small compared to the angular momentum
of all other objects in the system. This property moderates the degree of tilt misalignment produced by the resonance
considered in this paper, but also broadens its range of influence.

\cite{XiangGruess2014} considered the evolution of a tilted disk with
a planet in a binary and found approximate planet--disk coplanarity was maintained over the {$ \sim10$ binary orbital periods}
that they simulated.  Their system has an initial disk mass  of $5 \times 10^{-3} M$ (where $M$ is the binary mass)
that comes closest to our model with disk mass  $4 \times 10^{-3} M$  at $t=0$ 
seen in Figs.~\ref{planetdisc} and \ref{sph4} that shows significant misalignment over that time interval. 
Even higher disk mass models than considered by  \cite{XiangGruess2014}  show substantial misalignment
over this timescale (see Figs.~\ref{deltaimass} and \ref{sph2}). In particular, Fig.~\ref{deltaimass} shows that
significant misalignment occurs to disk masses of $0.01 M$.
Their simulated disk does not have a fully cleared gap, but instead
has a partial density depression at the orbit of the planet, as is seen in their Figs. 2 and 7.  Our simulations 
begin with a planet fully embedded in a disk without a gap.
During an initial adjustment phase, the disk evolves to have a clear gap near the orbit of the planet, as seen here in
Fig.~\ref{sigmaev}.
For the parameters of this model, the standard gap opening criterion is satisfied \citep{LP1986}.
As shown
in Section \ref{sec:gap}, substantial misalignment does not occur  for a system without a significant gap. \cite{Picogna2015} also found in SPH simulations that substantial misalignment occurs within
planet-disk systems.  Their simulations involve initial tilts relative to binary orbital plane
of $45^\circ$ and $60^\circ$, as were also simulated
by \cite{XiangGruess2014}.

We have assumed that the binary orbit is circular,
while binaries typically have eccentricities $\sim 0.4$. The secular model described in Section
\ref{sec:seceq} could be adapted to account
for the binary eccentricity by modifying the
coupling coefficients, while assuming the planet orbit and disk remain circular. 
Another effect of binary eccentricity is to
truncate the nearly coplanar disk to a smaller radius in terms of
binary semi-major axis $a$ \citep{Artymowicz1994}. 
For moderate eccentricity, we may
expect qualitatively similar behavior to the circular
case, but do not pursue the analysis in this paper.


The results of this work have several implications for giant planet
gas accretion. If a planet forms in a massive misaligned disk, it will
remain coplanar with the precessing disk until it becomes massive
enough to open a gap in the disk. Once a gap opens, the torque from
the disk on the planet becomes weaker. Thus, the disk and the
planet orbit may not remain coplanar.  The planet--disk misalignment 
is enhanced by the effects of the secular resonance
and also the effects of disk dissipation that cause disk tilt decay towards the binary orbital plane.
If the planet gains substantial mass from the disk after its tilt has decayed,
the planet will become more aligned with the disk and binary.
Also, if terrestrial planets form at a late stage from the remains
of the disk, then they may not be aligned to  giant
planet orbit. Instead they are likely to be more closely aligned to the binary orbital plane.


\section{Summary}
\label{sum}

We have explored the evolution of  planet--disk systems that orbit
a member of a binary star system. The planet, disk, and binary interact
through gravitational forces. The planet is taken to have an initial mass  of 0.1\% of the binary mass
that is large enough mass to open a gap in the disk. The planet orbit and disk
are initially coplanar and mildly inclined ($\sim 10^\circ$) relative to the orbital planet of the binary. 

The planet--disk system undergoes secular oscillations. Over the course
of the oscillations,
the planet and disk generally have 
a level of misalignment that is comparable to the initial planet--disk tilt relative to the binary
orbital plane for the parameters we considered.  The misalignment is aided by the effects of a secular resonance and the decay
of the disk tilt to the binary orbit plane.   At later times, the planet orbit can evolve towards alignment
with the disk, if the planet has gained a substantial amount of mass from a disk that has become
nearly aligned with the binary orbital plane. This tendency toward alignment is due to the advection of disk momentum  by
the planet.

We determined
 the  tilt evolution of the planet and disk by means of secular
theory and SPH simulations.  The secular model describes the general
properties of the gravitational interactions between the planet, disk, and binary
that are found SPH simulations. 
Since the secular model parameters were not
tuned to match the simulations,
the quantitative agreement is approximate.

In Section \ref{secular}, we apply the secular theory to a planet--disk system in a binary
for a nondissipative
disk that lies external to the orbit a planet with a clearance that depends
on the gap size. 
The tilts of the planet and disk undergo oscillations as the objects precess.
One would expect that at very small disk mass,
the disk and planet precess independently 
with substantial misalignment between the orbital plane of the planet relative to the disk plane. 
In addition one expects that for very high disk mass, the planet precession becomes locked to that of the
disk with a smaller relative tilt.
These expectations are realized in the secular model (see Fig.~\ref{deltaimass}). 

However, the amplitude of the relative planet--disk tilt
oscillations does not vary monotonically with disk mass. As seen in Figs.~\ref{deltaimass} and \ref{phasepor},
for small disk masses, the relative planet-disk tilt oscillation amplitude increases with disk mass
and reaches a peak value at a disk mass for which the planet--disk interactions are just strong enough
for them to begin to precess together in a mean sense (librate).  That is, precession rate locking (in a mean sense, i.e., libration)
 then does not guarantee coplanarity. Just the opposite occurs. When mean precession rate locking sets in as a function of increasing disk mass, the planet--disk misalignment is largest. We attribute this effect to a secular nodal resonance driven by the disk and binary companion as described in Section 
 \ref{sec:secres}.
This peak planet--disk relative tilt occurs at disk
masses that are several times the planet mass. The resonance is broad and enhances the misalignment
for higher disk masses. Substantial relative inclinations
between the planet orbit and the disk, of order the initial planet--disk tilts relative to the binary orbital  plane, 
are possible for outer disk masses $\ga 1\%$ of the binary mass.

By means of SPH simulations, we analyzed this process with a dissipative (viscous) disk.
The planet has an initial mass of $0.1\%$ of the binary mass. It advects mass and momentum from the disk and can migrate. The planet is initially embedded in a disk without gap. We describe the evolution after an
initial disk adjustment period of $10 P_{\rm b}$ discussed in Section \ref{sec:sph}. 
Several aspects of the SPH simulations agree with the secular model.
However, there are some differences.
The disk dissipation causes a decay of the disk tilt to the binary orbital plane. Consequently, the disk tilt does not rise back to its initial
value as occurs in the oscillations of the secular model.
Advection of disk momentum by the planet from a sufficiently large disk may cause the planet's orbit to evolve towards alignment at later times,
if it has gained substantial mass from a disk that has become nearly aligned with the binary orbital plane. 
At a higher disk mass, 0.6\% of the binary mass, the planet--disk system  evolves from nodal libration to circulation
as the disk mass decreases (Fig.~\ref{sph2}),  as discussed in Section~\ref{sec:sph}.

In the simulations, there is generally  a substantial
misalignment between the orbit of the planet and the disk that is comparable to the
initial tilt of the system (relative to the binary orbital plane).  The peak planet--disk misaligment occurs for a disk mass that is about 5 times the planet mass and significant  misalignment extends to higher disk masses, largely due to the effects of the secular resonance.  \cite{Picogna2015} also found in SPH simulations that substantial misalignment occurs within
planet-disk systems that orbit a member of a binary.


In this work, we have considered only mild initial misalignments between
the planet--disk system and the binary orbital plane. Recently, we found
that substantially misaligned disks (tilts between about $45^\circ$ and $135^\circ$ with respect to the binary orbital plane) can undergo coherent Kozai--Lidov tilt and eccentricity
oscillations \citep{Martinetal2014,Martinetal2014b, Fu2015}. 
 In a future paper, we will investigate
the evolution of planet--disk systems with such initial misalignments.


\acknowledgements
We thank the referee for useful comments and informing us about the paper by \cite{Picogna2015}.
We thank Jim Pringle for helpful conversations. SHL acknowledges support from NASA grant NNX11AK61G.  We acknowledge the use of SPLASH \citep{Price2007} for the rendering of the figures. Computer support was provided by UNLV's National Supercomputing Center. This work also utilised the Janus supercomputer, which is supported by the National Science Foundation (award number CNS--0821794), the University of Colorado Boulder, the University of Colorado Denver, and the National Center for Atmospheric Research. The Janus supercomputer is operated by the University of Colorado Boulder.



\label{lastpage}
\end{document}